# Glycolytic pyruvate kinase moonlighting activities

# in DNA replication initiation and elongation

Steff Horemans[1], Matthaios Pitoulias[2], Alexandria Holland[2], Panos Soultanas[2¶] and Laurent Janniere[1¶]

[1] : Génomique Métabolique, Genoscope, Institut François Jacob, CEA, CNRS, Univ Evry, Université Paris-Saclay, 91057 Evry, France

[2] : Biodiscovery Institute, School of Chemistry, University of Nottingham, University Park, Nottingham NG7 2RD, UK

Short title: PykA moonlighting activity in DNA replication

[¶] : Corresponding authors

Laurent Janniere: laurent.janniere@univ-evry.fr

Panos Soultanas : panos.soultanas@nottingham.ac.uk

**SUMMARY**

Cells have evolved a metabolic control of DNA replication to respond to a wide range of nutritional conditions. Accumulating data suggest that this poorly understood control depends, at least in part, on Central Carbon Metabolism (CCM). In *Bacillus subtilis,* the glycolytic pyruvate kinase (PykA) is intricately linked to replication. This 585 amino-acid-long enzyme comprises a catalytic (Cat) domain that binds to phosphoenolpyruvate (PEP) and ADP to produce pyruvate and ATP, and a C-terminal domain of unknown function. Interestingly, the C-terminal domain termed PEPut interacts with Cat and is homologous to a domain that, in other metabolic enzymes, is phosphorylated at a conserved TSH motif at the expense of PEP and ATP to drive sugar import and catalytic or regulatory activities. To gain insights into the role of PykA in replication, DNA synthesis was analyzed in various Cat and PEPut mutants grown in a medium where the metabolic activity of PykA is dispensable for growth. Measurements of replication parameters (*ori/ter* ratio, C period and fork speed) and of the pyruvate kinase activity showed that PykA mutants exhibit replication defects resulting from side chain modifications in the PykA protein rather than from a reduction of its metabolic activity. Interestingly, Cat and PEPut have distinct commitments in replication: while Cat impacts positively and negatively replication fork speed, PEPut stimulates initiation through a process depending on Cat-PEPut interaction and growth conditions. Residues binding to PEP and ADP in Cat, stabilizing the Cat-PEPut interaction and belonging to the TSH motif of PEPut were found important for the commitment of PykA in replication. *In vitro*, PykA affects the activities of replication enzymes (the polymerase DnaE, helicase DnaC and primase DnaG) essential for initiation and elongation and genetically linked to *pykA*. Our results thus connect replication initiation and elongation to CCM metabolites (PEP, ATP and ADP), critical Cat and PEPut residues and to multiple links between PykA and the replication enzymes DnaE, DnaC and DnaG. We propose that PykA is endowed with a moonlighting activity that senses the concentration of signaling metabolites and interacts with replication enzymes to convey information on the cellular metabolic state to the replication machinery and adjust replication initiation and elongation to metabolism. This defines a new type of replication regulator proposed to be part of the metabolic control that gates replication in the cell cycle.

## INTRODUCTION

Chromosome replication occurs once in every cell cycle in response to overlapping regulatory mechanisms that control the activity of replication initiators and replication origins. Chromosome replication is also coupled to the growth rate afforded by available nutrients. This nutrient-mediated growth rate regulation of DNA replication, termed metabolic control of replication, modulates the initiation frequency and/or speed of replication forks in bacteria (Bipatnath et al., 1998; Helmstetter, 1996; Schaechter et al., 1958; Sharpe et al., 1998). The net result is an precise and reproducible timing of DNA synthesis in the cell cycle, across a wide range of nutritional conditions and growth rates. In eukaryotes, the metabolic control of replication regulates S phase entry and progression, and confines DNA synthesis to the reduction phase of a redox metabolic cycle repeated several times per cell cycle (Buchakjian and Kornbluth, 2010; Burnetti et al., 2015; Ewald, 2018; Klevecz et al., 2004; Papagiannakis et al., 2017; Tu et al., 2005; Yu et al., 2009).

The mechanism of metabolic control of replication remains mostly elusive. In bacteria, long-standing hypotheses postulating that this control operates by modulating the concentration of the active form of the replication initiator DnaA (DnaA-ATP) or by restricting DNA polymerase activity through a limitation of precursor concentrations or by modulating the concentration of (p)ppGpp, an alarmon that accumulates under nutritional stress to inhibit replication initiation or elongation, have been challenged (Flåtten et al., 2015; Hernandez and Bremer, 1993; Hu et al., 2019; Mathews, 2015; Maya-Mendoza et al., 2018; Murray and Koh, 2014). It is also unlikely that the metabolic control operates by modulating the activity of DnaA and *oriC* regulators, as replication still responds to metabolism in cells lacking such regulators (Ishida et al., 2004; Lu et al., 1994; Murray and Koh, 2014). Hence, several groups have started to argue that the metabolic control of replication is a multifactorial process which varies with nutrient richness and may involve sensing cellular metabolism and communicating this information to the replication machinery (see for instance (Baranska et al., 2013; Boye and Nordström, 2003; Buchakjian and Kornbluth, 2010; Du and Stillman, 2002; Ewald, 2018; Wang and Levin, 2009; Yu et al., 2009)).

Nutrient richness is precisely sensed by central carbon metabolism (CCM), a group of reactions that break down carbon and nitrogen sources to produce the energy and precursors needed for cellular activities. These sensing and feeding activities make CCM a strategic hub for generating signals that cells may use to couple replication to cellular metabolism. Accordingly, an impressive number of CCM-replication links has been discovered from bacteria to higher eukaryotes. In the model bacterium *Escherichia coli*, these links mainly target the replication initiator DnaA and may thus adjust initiation to CCM activity. They comprise (i) a metabolite that gears CCM activity (cyclic AMP) interacts with DnaA to stimulate its binding to the replication origin and facilitate DnaA rejuvenation from the inactive DnaA-ADP to the active DnaA-ATP form (Hughes et al., 1988); (ii) changes in the pyruvate-acetate node suppress initiation defects of a DnaA mutant (*dnaA46*) (Maciag-Dorszynska et al., 2012; Maciąg et al., 2011; Tymecka-Mulik et al., 2017); and (iii) two metabolites of the pyruvate-acetate node (acetyl-CoA and acetyl-phosphate) drive DnaA acetylation to prevent DnaA from binding to ATP and *oriC* (Zhang et al., 2016). In addition to targeting initiation, *E. coli* CCM-replication links may also influence elongation (Krause et al., 2020; Maciag-Dorszynska et al., 2012; Maciąg et al., 2011).

CCM-replication links have also been discovered in *Bacillus subtilis*. First, subunits of the pyruvate dehydrogenase (PdhC) and related enzymes were shown to inhibit replication initiation and/or bind to the origin region or to replication enzymes (the helicase DnaC and the primase DnaG) (Laffan and Firshein, 1987; 1988; Noirot-Gros et al., 2002; Stein and Firshein, 2000). Second, CCM mutations were shown to suppress initiation and elongation defects in mutants of DnaC, DnaG and the lagging strand polymerase DnaE and to disturb replication initiation and elongation in a medium-dependent manner (Jannière et al., 2007; Nouri et al., 2018; Paschalis et al., 2017). Third, the metabolic control of replication is disturbed in cells mutated for glyceraldehyde 3-phosphate dehydrogenase (GapA), pdhB and the pyruvate kinase (PykA) (Murray and Koh, 2014). Collectively, these results support the signaling hypothesis mentioned above and further suggest that metabolic signals coupling replication to growth in *B. subtilis* originate from the CCM area that converts glyceraldehyde 3-phosphate to acetate (thick bars Fig. 1A) and impact at least three replication enzymes involved in replication initiation and elongation: DnaC, DnaG and DnaE (Dervyn et al., 2001; Le Chatelier et al., 2004; Paschalis et al., 2017; Rannou et al., 2013; Sanders et al., 2010; Soultanas, 2012). Although the underlying mechanism remains elusive, it was proposed that in some growth conditions, CCM modulates initiation by altering the functional recruitment of DnaC, DnaG and DnaE to *oriC*, and elongation by altering the activity of the lagging strand polymerase DnaE in replication forks (Nouri et al., 2018).

CCM-replication links were also found in eukaryotes ((Dickinson and Williams, 1987; Fornalewicz et al., 2017; Konieczna et al., 2015a; Sprague, 1977; Wieczorek et al., 2018) and see below). At first, these findings are surprising as metabolism and replication occur in different cellular compartments. However, an accumulation of data since the late 1950's shows that this dogma has exceptions, with numerous CCM enzymes being present in the nucleus. In this compartment, CCM enzymes were shown to produce metabolites impermeable to membranes (like ATP, acetyl-CoA, NADH) and to ensure non-metabolic (moonlighting) functions (see for instance (Boukouris et al., 2016; Kim and Dang, 2005; Konieczna et al., 2015b; Lu and Hunter, 2018; Ronai, 1993; Sirover, 1999; 2011; Snaebjornsson and Schulze, 2018)). Hence, CCM-replication links in eukaryotes may involve nuclear CCM determinants. Several data support this notion. For instance, the time of replication origin firing in the eukaryotic cell cycle depends on an increase in acetyl-CoA that promotes histone acetylation. This increase is geared by the redox metabolic cycle in yeast or by nuclear forms of the ATP-citrate lyase and Pdh complexes in mammalian cells (Cai et al., 2011; Sutendra et al., 2014; Wellen et al., 2009). Moreover, the nuclear form of GAPDH and lactate dehydrogenase are part of a cofactor of the transcription factor Oct-1, stimulates S phase progression by inducing expression of histone H2B in S phase (Dai et al., 2008; Zheng et al., 2003) while exogenous pyruvate represses histone gene expression and delays S phase entry (Ma et al., 2019). A nuclear form of phosphoglycerate kinase (PGK) interacts with the protein kinase CDC7 and positively regulates replication initiation by keeping the stimulatory effect of CDC7 on the MCM helicase (Li et al., 2018). Finally, some nuclear CCM enzymes (PGK, GAPDH and lactate dehydrogenase) modulate the activity of eukaryotic replicative polymerases (Pol$\alpha$, Pol$\varepsilon$ and Pol$\delta$) *in vitro* (Grosse et al., 1986; Jindal and Vishwanatha, 1990; Popanda et al., 1998). Overall, these data highlight the prime importance of CCM in replication control from bacteria to human cells.

Here, we report on the relationship between PykA and DNA replication in *B. subtilis* (Jannière et al., 2007; Murray and Koh, 2014; Nouri et al., 2018). PykA is a highly conserved homotetramer that catalyzes the final reaction of glycolysis, converting PEP and ADP into pyruvate and ATP. Its 3D structure has been solved for many organisms (Schormann et al., 2019) and residues of the catalytic site important for PEP, ADP binding and for phosphoryl transfer have been identified (Fig. S1A).

The PykA proteins of *B. subtilis* and related species have an extra C-terminal sequence (residues 477-585) (Nguyen and Saier, 1995; Sakai, 2004). Its deletion has no effect on the metabolic activity of the purified *G. stearothermophilus* PykA and its biological function is still unknown (Sakai, 2004). The crystal structure of *G. stearothermophilus* PykA indicates that the extra sequence interacts with the catalytic domain (termed thereafter Cat) through a hydrogen bond between E209 and L536 of the Cat and extra C-terminal sequence, respectively (Suzuki et al., 2008). Interestingly, the sequence and 3D structure of the extra C-terminus are homologous to one of the PEP utilizer (PEPut) domains found in the EI component of the phosphoenolpyruvate:carbohydrate phosphotransferase system, the pyruvate phosphate dikinase (PPDK) and the PEP synthase (PEPS) (Fig. S1B) (Sakai, 2004; Suzuki et al., 2008). In EI, PEPut has a histidine residue that transfers a phosphoryl group from PEP to a protein during the process of sugar transport (Alpert et al., 1985; Teplyakov et al., 2006). By contrast, in PPDK and PEPS enzymes, this H residue is a key part of the catalytic site: it is essential for the reversible catalysis of pyruvate into PEP and the reaction depends on the transfer of a phosphoryl group from ATP or PEP to pyruvate or AMP through transient phosphorylation of the conserved histidine (Goss et al., 1980; Herzberg et al., 1996). Interestingly, a LTSH motif, including the Cat-PEPut interacting L536 residue and the catalytic H, is conserved in PEPuts (Fig. S1B) (Sakai, 2004). The T residue of this motif is phosphorylated for inhibiting the catalytic activity of H in PEPS and PPDK proteins and its phosphorylation/dephosphorylation is catalyzed by serine/threonine kinases of the DUF299 family using ADP as donor (Burnell and Hatch, 1984; Burnell, 2010; Burnell and Chastain, 2006; Tolentino et al., 2013). This residue can also be phosphorylated in the PykA of *Listeria monocytogenes* (Misra et al., 2011). In contrast, the S residue of the LTSH motif has no effect on catalytic activity or substrate specificity (Tolentino et al., 2013). However, this residue is phosphorylated *in vivo* in *B. subtilis* and *Arabidopsis thaliana* (Eymann et al., 2004; Mäder et al., 2012; Reiland et al., 2009).

Here, we demonstrate that PykA is important for replication independently of its metabolic activity and its mere effects on cellular metabolism and growth rate. Interestingly, this replication function appears to depend on the protein itself with the Cat domain impacting positively and/or negatively on the replication fork speed, and the PEPut domain stimulating replication initiation through a Cat-PEPut interaction. Residues important for these effects on replication are amino-acids binding to PEP and ADP, the TSH motif, and residues stabilizing the Cat-PEPut interaction. We also show that PykA modulates the activity of replication enzymes involved in replication initiation and elongation *in vitro*. We propose that our findings uncover a new type of replication regulator that informs the replication machinery on the cellular metabolic state. This lays the foundations for wider studies of the underlying mechanisms and the basis for the metabolic control of replication.

## RESULTS

### 1. DNA replication defects in *pykA* null cells grown in a medium where *pykA* is dispensable for growth

Replication phenotypes in *ΔpykA* cells were found in conditions where metabolism and growth are dramatically reduced (Jannière et al., 2007; Murray and Koh, 2014; Nouri et al., 2018). To determine whether these reductions are mandatory for replication phenotypes to occur in the *ΔpykA* context, a study was carried out in a medium where *pykA* was dispensable for growth (MC medium, Fig. 1B). Under this nutritional condition, replication phenotypes were still observed in *pykA* knockout cells: in comparison to the wild-type strain, the mutant has a higher *ori/ter* ratio (the relative copy number of origin versus terminus sequences), an extended C period (the time required for replicating the entire chromosome), a lower speed of replication forks and a lower number of origins per cell (Fig. 1C-E and Table S1). This shows that replication defects in *ΔpykA* cells depend on a reduced PykA activity (this activity is reduced ~25-fold in the mutant in comparison to the wild-type strain (Fig. 1B)) and/or the absence of the protein itself. However, these phenotypes do not depend on a mere decrease in growth and metabolism. Similar results were found when the ability of PykA depletion to suppress thermosensitive mutations in replication enzymes was assessed (Jannière et al., 2007).

## 2. Residues of both the Cat and PEPut domains of PykA impact replication through pathways independent of the PykA catalytic activity

*Replication analysis in Cat mutants*

To further investigate the involvement of PykA in replication, we next analyzed the *ori/ter* ratio, the C period and the replication fork speed in PykA catalytic mutants cultured in the MC medium. In this study, we used a strain deleted for the catalytic domain (Cat; $pykA_{\Delta cat}$) and mutants that affect the binding to PEP and ADP substrates ($pykA_{R32A}$, $pykA_{R73A}$, $pykA_{GD245\text{-}6AA}$ and $pykA_{T278A}$) or facilitate the phosphoryl transfer during catalysis ($pykA_{K220A}$) (Fig. S1A) (Bollenbach et al., 1999; Schormann et al., 2019). We also used a 27 amino-acid deletion (208-234) ($pykA_{JP}$) that removes the residues involved in the stabilization of the Cat-PEPut interaction (E209) and the phosphoryl transfer helper (K220). This deletion uncovered a genetic link between PykA and the lagging strand polymerase DnaE (Jannière et al., 2007).

Four classes of mutants characterized by a low, wild-type, high and notably high *ori/ter* ratio were identified using careful measurements of the ratio and the Mann-Whitney statistical test run at a significance $P < 0.01$ (Fig. 2A, Table S1). These mutants generally exhibited additional replication phenotypes which tend to covary (positively for the C period and negatively for the fork speed) with the *ori/ter* ratio (Fig. 2B). The PykA activity in crude extracts of Cat mutants was reduced to the same extent as in the *ΔpykA* strain (Fig. 2A), showing that Cat mutations dramatically inhibit (or abolish) PykA metabolic activity, as expected. Growth curve analysis in a glycolytic medium (GC) confirmed PykA inhibition and further showed that Cat mutations, like *pykA* deletion, do not affect growth in the MC medium (Fig. S2A). Overall, our data show that the Cat domain of PykA impacts replication in the MC medium (where catalysis by PykA is not needed) through different pathways, with side chain modifications to residues critical for PykA metabolic activity associated with no or opposing replication phenotypes. Since the Cat mutants have different replication defects and a similar residual metabolic activity, these results also suggest that the Cat-replication relationship depends on amino-acid side chains of the PykA protein rather than on its metabolic activity.

*Replication analysis in PEPut mutants*

Next, we analyzed replication parameters in PEPut mutants grown in the MC medium. This study included a strain encoding a PykA protein deleted for PEPut ($pykA_{\Delta PEPut}$) and mutants of the TSH motif in which individual amino-acids, or the whole motif, were replaced by A or D, the latter amino-acid mimicking phosphorylation. We also tested mutants of the Cat-PEPut interaction (*pykA*$_{E209A}$ and *pykA*$_{L536A}$) (Suzuki et al., 2008).

Three classes of PEPut mutants characterized by a very low, low and wild-type *ori/ter* ratios were identified by the Mann-Whitney test run at a significance $P < 0.01$ (Fig. 3A, Table S1). Mutants with a very low ratio (T>D and TSH>DDD) exhibited additional replication phenotypes: a remarkably short C period and a very high fork velocity (Fig. 3B). In contrast, mutants with a low ratio had no clear extra phenotypes. We speculate that this mitigated response results from the fact that the primary target of PEPut is initiation (see below) while the C period and fork speed mainly probe elongation defects.

The S residue of the TSH motif and the amino-acids stabilizing the Cat-PEPut interaction (E209 and L536) are important for replication in the MC medium, as the corresponding A mutants exhibited replication phenotypes. In contrast, the lack of phenotype in T>A and H>A cells shows that these residues are dispensable for replication in this medium. However, in growth conditions favoring phosphorylation, T and H become important for replication, as clear phenotypes were found in T>D and H>D cells.

Surprisingly, PykA dosage in crude extracts showed that PEPut mutations have strong and opposite effects on the metabolic activity of PykA. While this activity is increased (+ 25 to 50 %) in the T>A, S>A and H>A mutants, it is reduced (- 50 %) in cells deleted for PEPut or impeded in the PEPut-Cat interaction, and dramatically inhibited (- 75-90 %) in the TSH>AAA, TSH>DDD and T>D contexts (Fig. 3A). In contrast, this activity remains unchanged in H>D and S>D cells. We also found that changes in the PykA activity and the *ori/ter* ratio do not covary in PEPut mutants (Fig. 3A) and that these mutants have a wild-type growth rate in the MC medium (Fig. S2B). It is inferred from these data that, as for Cat mutants, changes in PykA catalytic activity cannot account for replication defects in PEPut mutants. Overall, our findings show that the PEPut-replication relationship depends on the TSH motif, its phosphorylation status and on residues stabilizing the Cat-PEPut interaction. It however does not depend on the metabolic activity of PykA and/or a mere reduction in cellular metabolism. Interestingly, our results assign a strong (up to 150-fold) regulatory function to PEPut on PykA activity and showed that this regulatory activity has the same requirements as the PEPut-replication relationship: the Cat-PEPut interaction, the TSH motif and the level of TSH phosphorylation. We anticipate that this regulatory activity is driven by metabolic signals that adjust the concentration of PykA products in a wide range of nutritional conditions. This regulation provides an elegant solution to the problem posed by a constitutive and abundant production of PykA (Mäder et al., 2012; Nicolas et al., 2012).

**3. The replication phase targeted by Cat and PEPut are elongation and initiation, respectively**

Changes in replication initiation are often compensated by opposite changes in elongation and *vice versa* (see (Morigen et al., 2009; Odsbu et al., 2009; Skarstad et al., 1989). This precludes the identification of the initial target (initiation or elongation) of *pykA* mutations. To clarify this issue, we constructed a series of strains in which replication is initiated from a plasmid replicon (*oriN*) integrated into the chromosome, close to *oriC*; this replication is independently of the chromosomal initiation factors DnaA and *oriC* and uses a chromosomal-type replication fork for copying DNA (Hassan et al., 1997). As argued previously (Murray and Koh, 2014; Nouri et al., 2018), if replication defects in *pykA* mutants were to result from changes in initiation (i.e. the productive interaction of DnaA with *oriC*), cells replicating their genome from the plasmid replicon would not suffer from the metabolic mutation and would thus exhibit the *ori/ter* ratio of PykA+ *oriN*-dependent cells. In contrast, if the replication defects were to result from changes in elongation (or in an initiation event downstream of the formation of an active DnaA/*oriC* complex), plasmid replicon-dependent cells would still suffer from the *pykA* mutation and would thus have an *ori/ter* ratio different from PykA+ *oriN*-dependent cells.

The analysis was carried out in the $\Delta$*oriC* context, rather than in the $\Delta$*dnaA* context, to avoid any interference of DnaA depletion on genome expression (Washington et al., 2017). The tested mutations were representatives of the different classes of Cat ($pykA_{T278A}$, $pykA_{JP}$, $pykA_{GD256/6AA}$) and PEPut ($pykA_{T>D}$ and $pykA_{TSH>AAA}$) mutants (Fig. 2 and 3). Mutations affecting the Cat-PEPut interaction ($pykA_{E209A}$ and $pykA_{L536A}$) were included in the study. Results showed a ratio typical to PykA+ *oriN*-dependent cells in PEPut and Cat-PEPut interaction mutants and a clearly reduced ($pykA_{JP}$, $pykA_{GD256/6AA}$) or increased ($pykA_{T278A}$) ratio in Cat mutants (Fig. 4). Hence, Cat and PEPut have different primary replication targets: initiation (formerly the DnaA/*oriC* interaction) for PEPut and the Cat-PEPut interaction, and elongation (or an initiation step downstream of the formation of an active DnaA/*oriC* complex) for Cat. Moreover, PEPut acts as an activator of initiation that operates via its interaction with Cat while Cat acts as a positive and/or negative effector of elongation. Overall, these results show that the role of PykA has multiple roles in replication that involves distinct Cat and PEPut non-metabolic functions, as well as the Cat-PEPut interaction.

**4. The purified PykA protein modulates the replication activities of DnaE and DnaC**

To investigate whether PykA impacts replication by directly modulating the activity of replication enzymes, the effect of PykA on replication enzymes' activities was analyzed *in vitro*. The PykA protein was heterologously expressed and purified, and control experiments showed that the purified PykA was metabolically active and formed the expected stable functional tetramer (Fig. S3). The effect of PykA was then tested on DnaC, DnaG and DnaE, as these replication enzymes are genetically linked to PykA (Jannière et al., 2007). The DnaC helicase melts the duplex DNA at *oriC* during initiation and separates the DNA strands in replication forks during elongation. The DnaG primase synthesizes RNA primers at the origin and in replication forks. These primers are extended by DnaE to start leading strand synthesis (which is mainly carried out by PolC) and to ensure partial or complete synthesis of the lagging strand. Previous studies showed that DnaC, DnaG and DnaE form a ternary complex that ensures important roles during replication initiation and elongation (Dervyn et al., 2001; Paschalis et al., 2017; Rannou et al., 2013; Sanders et al., 2010).The three replication enzymes were purified and their activities were tested as described previously (Paschalis et al., 2017; Rannou et al., 2013) in the presence or absence of PykA.

Replication assays with DnaE were carried out at a polymerase concentration (10 nM) that produces a low amount of replication products in order to facilitate the detection of stimulatory effects. In reaction mixtures containing M13 ssDNA annealed to a 20-mer DNA oligonucleotide and DnaE in combination or not with equimolar concentrations of PykA, a substantial stimulation of DnaE polymerase activity by PykA was found in terms of both sizes and amounts of nascent DNA synthesized (Fig. 5A, left panel). A similar stimulation was observed with a 60-mer oligonucleotide annealed onto M13 ssDNA and a 15-mer oligonucleotide annealed onto a 110-mer oligonucleotide (Fig. S4A and data not shown). The stimulation was specific to PykA as it was not observed with equimolar amounts of BSA and a M13 ssDNA primed with a 20-mer (Fig. 5A, right panel) or a 60-mer (Fig. S4B). The lack of DnaE stimulation by BSA was further confirmed at a 50-fold excess concentration over DnaE with the 20-mer primed M13 ssDNA (Fig. S4C) (note that the marginal stimulation observed at very high (500-fold) BSA excess is artifactual, acting likely as a blocking agent preventing adhesion of DnaE to the plastic reaction tubes). Titration experiments and gel shift assays showed that the stimulation of DNA synthesis by PykA was not due to a stimulation of DnaE binding to primed template (Fig. S5).

Within the replisome, the polymerase activity of DnaE is stimulated by several proteins (Bruck and O'Donnell, 2000; Le Chatelier et al., 2004; Paschalis et al., 2017; Rannou et al., 2013), with DnaN being a potent stimulator. This ring-shaped protein is loaded at the 3' end of primed sites by the clamp loader, encircles the DNA and binds DnaE to form a stable complex that slides along the DNA template. In order to determine whether the polymerase activity of the DnaE-DnaN complex can be further stimulated by PykA, we carried out primer extension assays with DnaE, DnaN, proteins of the clamp loader (HolB, YqeN and DnaX) and PykA (Fig. 5B). As previously observed (Paschalis et al., 2017), we found that the polymerase activity of DnaE (10 nM) is strongly stimulated by DnaN and the clamp loader (compare the left panels of Fig. 5A and B). In this condition of high activity, the effect of PykA is unclear (Fig. 5B). However, this glycolytic enzyme may confer an additional stimulation to the DnaE activity as its presence leads at the last timepoints to a greater accumulation of large fragments and a clear reduction in labelled primer (Fig. 5B, compare with and without PykA). To confirm this, we carried out similar primer extension assays in the presence of lower, suboptimal concentrations of DnaE (2 nM), DnaN, clamp loader proteins and PykA (in both sets of assay conditions, the molar ratios of proteins were kept identical). At this suboptimal DnaE concentration, no nascent DNA was detectable in the absence of PykA and significant amounts of nascent DNA fragments were synthesized by DnaE in the presence of PykA (Fig. 5C). This suggests that PykA stimulates the activity of DnaE even in conditions where its activity is strongly stimulated by DnaN. It is inferred from these data that PykA stimulates the DnaE polymerase activity when the polymerase is slow (i.e. alone) and fast (i.e. in a complex with DnaN) probably via a direct interaction between PykA and DnaE. Since the purified PEPut domain does not affect DnaE activity (Fig. S6-7), the stimulation probably depends on a direct interaction between DnaE and the Cat domain of PykA or the interaction interface involves structural features of the PykA tetramer that are not preserved in the purified PEPut. The stimulation occurs with short (20-mer) and long (60-mer) primers suggesting that PykA may stimulate DnaE polymerase activity during extension of RNA primers generated by DnaG and during lagging strand synthesis.

The helicase activity of DnaC was assayed by monitoring the displacement of a labelled 104-mer oligonucleotide annealed onto M13 ssDNA and forming a double-fork substrate with poly(dT) tails as previously (Rannou et al., 2013). To assemble a functional DnaC hexamer onto the DNA substrate, the helicase loader DnaI was added to reaction mixtures at equimolar concentrations. Results showed that DnaC is marginally inhibited in the presence of PykA (Fig. 6A). As we previously showed that DnaC activity is stimulated by DnaG (Rannou et al., 2013), DnaG was added to the above-mentioned reaction mixtures at equimolar concentrations. This analysis confirmed the stimulation of helicase activity by DnaG and showed that this stimulation is cancelled by PykA (Fig. 6B). Thus, in some contexts, PykA can significantly inhibit the helicase activity of DnaC. Collectively, these results suggest that PykA can modulate replication enzyme activities through direct functional interactions with replication enzymes.

## DISCUSSION

PykA is one of the most conserved and abundant proteins of the evolution tree and its ability to convert PEP and ADP into pyruvate and ATP during glycolysis is known in exquisite detail. However, the interest in this well-characterized protein has been rekindled in the last decades by various studies showing that PykA from bacteria to higher eukaryotes often mediates non-metabolic (moonlighting) functions in processes as diverse as transcription, viral replication, angiogenesis, pathogenicity and tumorigenesis (Boukouris et al., 2016; Chuang et al., 2017; Lu and Hunter, 2018; Pancholi and Chhatwal, 2003; Snaebjornsson and Schulze, 2018). Here, we provide evidence that PykA has a moonlighting role in *B. subtilis* chromosome replication.

Links between *pykA* and replication in *B. subtilis* were initially discovered in nutritional conditions where PykA is important for growth. Under these conditions, *pykA* deletion is associated with several replication phenotypes: (i) suppression of thermosensitive mutations in the replication enzymes DnaC, DnaG and DnaE but not DnaI, DnaD, PolC, DnaX and DnaN (Jannière et al., 2007), (ii) stimulation of initiation (Nouri et al., 2018), and (iii) alteration of the metabolic control of replication (Murray and Koh, 2014). Here we show that a growth rate decrease is not mandatory for replication defects to occur in *pykA* knock-out mutants, as several replication phenotypes were found in the MC medium where PykA was dispensable for growth (Fig. 1). It is interesting to note that, although dispensable, wild-type cells produce as much PykA protein in MC as in the LB medium (Fig. S8) where PykA is important for growth and replication (Jannière et al., 2007; Murray and Koh, 2014; Nouri et al., 2018).

To gain insights into the *pykA*-replication relationship, DNA synthesis was monitored in Cat and PEPut mutants grown in MC. Cat mutations affect highly conserved residues of the catalytic site while PEPut mutations affect the conserved TSH motif (see above). We also tested mutations assumed to destabilize the Cat-PEPut interaction ($pykA_{E209A}$ and $pykA_{L536A}$). Significant replication phenotypes were found in most of the 18 tested mutants (Fig. 2 and 3). These phenotypes are not due to changes in the PykA metabolic activity, as this activity does not covary with the *ori/ter* ratio, a global replication tracer accurately monitored in this study (Fig. 2 and 3). They are rather due to change in the PykA protein itself. This hypothesis is further supported by *in vitro* studies aimed at analyzing the effect of purified PykA on replication enzyme activities in conditions that do not permit PykA metabolic activity. These assays were focused on replication enzymes genetically linked to PykA, namely the helicase DnaC, the primase DnaG and the lagging strand polymerase DnaE (Jannière et al., 2007). Results showed that PykA stimulates the polymerase activity of DnaE likely via a direct PykA-DnaE interaction and inhibits the helicase activity of DnaC and the stimulatory effect of DnaG on DnaC (Fig. 5 and 6). Moreover, the effect of PykA on DnaE activity may occur in conditions where the polymerase is slow (alone) and fast (bound to DnaN) and during primer extension and lagging strand synthesis. Using *pykA* mutants replicating the genome from a plasmid origin, rather than from the chromosomal origin, we also found that the initial replication targets of the Cat and PEPut domains are different: elongation for Cat mutants and initiation for PEPut and Cat-PEPut interaction mutants (Fig. 4). Overall, our results suggest that the involvement of *pykA* in replication in the MC medium is an additional, non-metabolic, function of PykA. This moonlighting function appears to be divided into two components with the Cat domain either speeding up or slowing down the speed of replication fork, and PEPut stimulating replication initiation through a process depending on Cat-PEPut interaction and growth conditions. Furthermore, *in vitro* studies suggest that these moonlighting activities may involve interactions between PykA and replication enzymes.

Amino-acids that interact with PykA substrates (PEP and ADP;  R32, R73, D245-G246, T278) are important for Cat moonlighting activity in elongation. In contrast, this activity does not depend on the residue facilitating the phosphoryl transfer during catalysis (K220) (Fig. 2). This suggests that the impact of Cat on DNA elongation depends on PykA binding to its substrates. Moreover, as binding mutants are associated with a rather large range of fork velocities (from 600 to 870 bp/sec) (Fig. 2), the impact of Cat on elongation may be geared by the intracellular concentration of PykA-substrate complexes. Accordingly, the fork speed decrease (650 bp/sec) found in cells encoding a PykA protein truncated for residues 208-234 ($PykA_{JP}$, Fig. 2) may result from a defect in the formation of PykA-substrate complexes. Interestingly, this in-frame deletion was previously associated with changes in DnaE polymerase activity and conformation (Jannière et al., 2007). Hence, our results link fork speed to PykA-substrate concentration and DnaE (and possibly DnaC and DnaG) activity and conformation. As substrate binding causes conformational changes, we speculate that the Cat moonlighting activity in DNA elongation is regulated allosterically by conformational changes induced by PEP and ADP binding to PykA.

The LTSH region is important for PEPut moonlighting activity in initiation. In PykA, the L residue stabilizes an interaction between Cat and PEPut through a hydrogen bond with E209 (see above). Results show that this interaction is important for PEPut moonlighting activity as the L>A and E209A mutants have similar replication phenotypes: a low *ori/ter* ratio and an initiation defect (Fig. 3 & 4). The latter phenotype of the E209A mutant is remarkable as the main replication target of Cat is elongation (see above). The role of the conserved TSH motif in PEPut moonlighting activity appears to be medium-dependent (Fig. 3).  Individual amino acid residue replacements by A show that only the S amino acid of the motif is important for replication in the MC medium. However, the replacement of T and H by D, a phospho-mimicking amino-acid, is associated with replication defects. This suggests that T and H contribute to PEPut moonlighting activity in growth conditions where these residues are phosphorylated. In support of this hypothesis are data from bacteria to plants showing that the TSH motif is phosphorylated and that T and H modifications ensure catalytic and regulatory functions (see above). Collectively, our results show that the PEPut moonlighting activity in initiation stimulation depends on (i) the Cat-PEPut interaction and on (ii) the TSH motif which operates in a medium-dependent manner and possibly through phosphorylation. As T and H phosphorylations occur at the expense of PEP and ATP in PEPut containing metabolic enzymes (see above), these results may connect initiation to PEP and ATP concentration in growth conditions permissive for T and/or H phosphorylation.

In conclusion, our findings suggest that the Cat and PEPut domains of PykA sense the concentration of PEP, ATP and ADP to convey signals to DnaC, DnaG and DnaE on the metabolic state and modulate accordingly replication initiation and elongation. PykA may thus be a prototype for a new family of replication regulator that operates in the metabolic control of replication to assist replication gating in a wide range of growth conditions (see our model Fig. 7). We propose that the ever-increasing volume of data describing genetic, functional and/or biochemical links between CCM determinants and DNA replication from bacteria to eukarya (see above) all converge on an emerging ubiquitous and highly evolved system that assigns a replication timing function to CCM. Interestingly, a literature survey shows that pyruvate kinases regulate glycolysis and downstream pathways via allosteric mechanisms involving protein-metabolite interactions and metabolite-driven post-translational modifications (Mäder et al., 2012; Pisithkul et al., 2015; Prakasam et al., 2018; Schormann et al., 2019). Additionally, we found that PEPut is a strong regulator of PykA metabolic activity (Fig. 3). We thus suggest that the *B. subtilis* PykA protein is a master regulator that senses a small number of key signaling CCM metabolites so as to integrate cellular metabolism and DNA replication.

## MATERIAL AND METHODS

### *Strains and plasmids*

Strains and plasmids are listed in Supplementary Table S2. DNA sequences are available upon request.

*Construction of the pykA-tet strain (DGRM295)*:

In order to facilitate the construction of *pykA* mutants, we first inserted a *tet* gene immediately downstream from the transcription terminator of *pykA*. We first amplified, using the Q5 high-fidelity DNA polymerase (New England Biolabs, Evry, Fr), the *tet* gene of plasmid pTB19 (Oskam et al., 1991) and two chromosomal sequences flanking the site of insertion and containing the 3' end of the *pfk* gene on the one hand and the *ytzA* plus the 3' end of *ytvI* on the other hand. Next, we separated the PCR products from the parental genomic DNA by gel electrophoresis and purified the PCR fragments using the Monarch DNA Gel Extraction Kit (New England Biolabs, Evry, Fr). The purified fragments were then mixed and fused together using assembly PCR carried out with the Q5 high-fidelity DNA polymerase. This reaction depends on 20 bp sequences homologous to the 5' and 3' tails of the *tet* fragment that were added to the internal side of the chromosomal PCR fragments. Competent cells of a wild-type strain cured of prophages (TF8A) were then transformed with the assembly PCR product and double cross-over events integrating the *tet* gene downstream of *pykA* were selected on Tet containing plates (before plating, the *tet* gene was induced by incubating cells 1h at 37°C in the presence of 1.5 µg/mL Tet). A representative transformant was selected by PCR and sequencing and named DGRM295.

*Construction of Cat and PEPut mutants*:

To generate Cat and PEPut mutants, pairs of PCR reactions were performed using as template the genomic DNA of DGRM295 (*pykA-tet*). In each reaction, one external (i.e. in *pfk* or *ytvI*) primer and one mutagenic primer mapping in Cat or PEPut were used to generate PCR products with the desired *pykA* mutation at one end. PCR fragments were then assembled and the assembly products used to transformed TF8A competent cells, as described above (inactive pyruvate kinase mutants were selected on LB + Tet + Malate 0,4%). Three representative transformants were selected by sequencing for all constructions.

*Construction of pykA mutants replicating the chromosome from oriN*:

To construct *pykA* mutants replicating the chromosome from *oriN*, competent cells of a TF8A derivative carrying *oriN* and the *cat* gene at the *spoIIIJ* locus and deleted for *oriC* (DGRM589, (Nouri et al., 2018)) were transformed with genomic DNA of *pykA* mutants. Transformants were then selected on plates supplemented with appropriate antibiotic (and 0.4 % malate when the mutation inactivated PykA activity). Three representative transformants were generally selected. The presence of the *pykA* mutation was confirmed by sequencing while the presence of the *oriN-cat* structure and the *oriC* deletion were checked by measuring the size and sensitivity to *EcoRI* restriction of the corresponding PCR products.

*Construction of a strain encoding the PykA protein fused to the FLAG-tag*

To generate a strain encoding PykA C-terminally fused to the FLAG-tag (DGRM1098), pairs of PCR reactions were performed using as template the genomic DNA of DGRM295 (*pykA-tet*). In each reaction, one external (i.e. in *pfk* or *ytvI*) primer and a primer adding the FLAG-tag (DYKDDDDK) preceded by the linker (SGSG) to the C-terminus of PykA were used to generate PCR products. PCR fragments were then assembled and the assembly products used to transformed TF8A competent cells, as described above. A representative clone was selected by DNA sequencing.

***Growth conditions***

Routinely, *B. subtilis* cells were grown at 37 $^{o}$C in LB with or without antibiotics at the following concentrations: spectinomycin (Sp, 60 µg/mL); tetracycline (Tet, 5 µg/ml); chloramphenicol (Cm, 5 µg/mL); phleomycin (Phl, 10 µg/mL); kanamycin (Kan, 5 µg/mL). Malate (0.4 %) was added to cultures of *pykA* catalytic mutants. Replication parameters were determined in cells grown at 37°C in MC, a minimal medium ($K_2HPO_4$: 80 mM; $KH_2PO_4$: 44 mM; $(NH_4)_2SO_4$: 15 mM; $C_6H_5Na_3O_7$ $2H_20$: 3, 4 mM; $CaCl_2$: 50 mM; $MgSO_4$: 2 mM; FeIII citrate: 11 µg/mL; $MnCl_2$: 10 µM; $FeSO_4$: 1 µM; $FeCl_3$:4 µg/mL; Trp 50 µg/mL) supplemented with 0.2% enzymatic casein hydrolysate, 0.4 % malate and 0.01 % tryptophan. Glucose (0.4 %) was used instead of malate in the MG medium.

***Quantitative PCR***

For monitoring *ori/ter* ratios, 6-14 cultures inoculated routinely from three independent constructs were first grown overnight at 30°C in MC supplemented with antibiotic. Upon saturation, cultures were diluted 1000-fold in the same medium without antibiotic and growth at 37°C was carefully monitored using spectrophotometry. Samples for qPCR analysis were collected at low cell concentration ($OD_{600\ nm}$ = 0.06 to 0.15) to ensure that cell cycle parameters are determined in steady-state cells and are not affected by the approach to the stationary phase or by changes in medium composition. The genomic DNA was extracted as described previously (Nouri et al., 2018) or using the PureLink Genomic DNA mini kit (Invitrogen by Thermo Fisher Scientific, Courtaboeuf, Fr). Every qPCR reaction was carried out using two technical repeats of 4 serial dilutions. A non-replicating control DNA (stage II sporlets, (Magill and Setlow, 1992)) was analyzed simultaneously with the samples in about 1/4 of the qPCR plates. Reactions and amplifications were carried out as previously described in 1x SYBR qPCR Premix Ex Taq (Tli RNaseH Plus) (Ozyme, St Quentin en Yvelines, France) mix and on a Mastercycler® ep realplex (Eppendorf, Le Pecq, Fr) (Nouri et al., 2018). Ratios of *pykA* mutants were normalized using the mean of about 125 measures of the non-replicating control DNA (0.5885 +/- 0.006) and compared using the nonparametric Mann-Whitney U test (https://www.socscistatistics.com/tests/mannwhitney/default2.aspx) run at a significance level of 0.01 and a two-tailed hypothesis.

C periods were determined from three independent cultures and using 10 pairs of primers arranged regularly (from *oriC* to *terC*) along the right arm of the chromosome, as previously described (Nouri et al., 2018). Mean fork velocity was calculated using the C period (min) and the actual size of the TF8A genome (4,000,631 bp; this genome is deleted for prophages SPβ, PBSX and *skin*).

***Flow cytometry analysis***

Strains were grown as indicated in the previous section and at $OD_{600nm}$ = 0.1–0.15, chloramphenicol (200 µg/mL) was added to the cultures to impede replication initiation, cell division and allow completion of ongoing rounds of replication (Séror et al., 1994). After 4 hr of drug treatment, $10^8$ cells were fixed in filtered ethanol 70% and stored at 4°C. Stored cells were then washed twice in 1 mL of filtered Tris-buffered saline buffer (TBS 150) (20 mM Tris-HCl pH 7.5, 150 mM NaCl) and stained with Hoechst 33258 (1.5 µg/mL) for at least 30 min, as described elsewhere (Morigen et al., 2009). Flow cytometry analysis was performed using a MoFlow Astrios cell sorter (Beckman Coulter, Life Sciences) equipped with a 355 nm krypton laser and a 448/59 nm bandpass filter used to collect Hoechst 33258 fluorescence data. Data were analyzed with the Kaluza software (Beckman Coulter, Life Sciences). We counted 100,000 events. In most of the tested samples, DNA histograms show 2 main peaks with a $2^n$ distribution. The number of origins/cell was obtained from the proportion of cells with 1, 2, 4, 8 and 16 chromosomes.

***Pyruvate kinase activity measurement***

Cells (25 mL; $OD_{600nm}$ = 0.3) growing exponentially in MC were carefully collected by centrifugation (7 min; 4,922 RFC; room temperature) and resuspended in 75 µL of lysis buffer ($Na_2HPO_4$ $2H_2O$: 60mM; $NaH_2PO_4$ $4H_2O$: 4mM; KCl: 10 mM; $MgSO_4$ 7H2O: 1mM; DTT: 1mM; Lyzozyme: 0.1 mg/mL; DNase I: 40 U/mL). They were then incubated 20 min on ice, 5 min at 37 °C and 15 min at room temperature. Crude extracts were then collected by centrifugation (10 min; 14,000 rpm; 4 °C). The PykA activity was determined using the colorimetric/fluorometric assay kit K709-100 (CliniScience, Nanterre, Fr) and fluorescence (Ex/Em = 535/587 nm) was assessed using a ClarioStar apparatus (BMG Labtech, Champigny-sur-Marne, Fr). Protein concentration was monitored using the standard Bradford assay.

***Western blot***

DGRM1098 cells growing exponentially in the LB or MC medium in the presence of Tet were collected at $OD_{650nm}$ = 0.5 by centrifugation (4,500 rpm, 5 min, 4°C) and washed 3 times in PBS 1x (4°C). They were then resuspended in 250 µL of freshly prepared lysis solution (CelLytic B Cell Lysis Reagent 1x – Sigma -, Tris-HCl pH 7.5 (50 mM), PMSF (0.5 mM), cOmplet ULTRA Tablets, Mini, EASYpack, EDTA-free (8 mg/mL) – Roche -, Lysozyme (0.5 mg/mL) – Sigma). Upon incubation at room-temperature under gentle shaking for 15 min, samples were centrifuged (15,000 rpm, 20 min, 4°C) and supernatants were collected and kept at -20°C. To prepare samples for SDS-PAGE, 40 µL of supernatant containing 5 µg of proteins were mixed to 36 µL Laemmli Buffer 1x (Bio-Rad) and 4 µL β-mercaptoethanol BioUltra (Sigma) and immediately heated for 10 min at 90 °C. The SDS-PAGE gel electrophoresis, Coomassie staining and Western blotting were performed according to standard procedures. Primary antibody: Monoclonal Anti-FLAG M2 (Sigma); Secondary antibody: Goat anti-mouse IgG, HRP conjugate, species adsorbed (Sigma); Kit ECL Western Blotting Detection Reagents (Amersham).

***Protein Biochemistry***

*Replication enzymes:*

Replication enzymes DnaC, DnaG, DnaE, DnaN, HolB, YqeN and DnaX were purified and tested as described previously (Paschalis et al., 2017; Rannou et al., 2013).

*Helicase assays:*

Briefly, DnaC helicase assays were carried out with 633 nM (referring to monomer) each DnaC and DnaI, in 50 mM NaCl, 50 mM Tris-HCl pH 7.5, 10 mM $MgCl_2$, 1 mM DTT, 2.5 mM ATP and 2 nM DNA substrate, in the presence and absence of DnaG (313 nM) and/or PykA (633 nM monomer). The DNA substrate was constructed by annealing a 5’-$^{32}$P radioactively labelled 100mer oligonucleotide (5’-CACACACACACACACACACACACACACACACACACACACACACACACACACACACACCCCTTTAAAAAAAAA AAAAAAAGCCAAAAGCAGTGCCAAGCTTGCATGCC-3’) onto M13 ssDNA. The helicase activity was assayed by monitoring the displacement of the radioactive oligonucleotide from the M13 ssDNA through non-denaturing PAGE using 10% w/v polyacrylamide gels. Data were quantified using a Personal Molecular Imager with Quantity One 1-D analysis software (Bio-Rad) and analyzed with GraphPad Prism 4 software.

*Polymerase assays:*

Time course polymerase assays were carried out by monitoring the DnaE primer extension activity using a 5'-$^{32}$P radioactively labelled 20mer (5'-CAGTGCCAAGCTTGCATGCC-3') or 60mer (5'-CAGTGCCAAGCTTGCATGCCTGCAGGTCGACTCTAGAGGATCCCCGGGTACCGAGCTCGA-3') annealed to M13 ssDNA substrates (2 nM), in 50 mM Tris-HCl pH7.5, 50 mM NaCl, 10 mM $MgCl_2$, 1 mM DTT, 1 mM dNTPs and 10 nM DnaE in the presence or absence of 40 nM PykA tetramer or 40 nM BSA. In some reactions with a lower, suboptimal 2 nM DnaE, 8 nM PykA tetramer was used. Nascent DNA was resolved through alkaline gel electrophoresis, as described before (Paschalis et al., 2017; Rannou et al., 2013). Visualisation and quantification were carried out using a Personal Molecular Imager with Quantity One 1-D analysis software (Bio-Rad) and data were analyzed with GraphPad Prism 4 software.

The effect of purified PEPut on the DnaE activity was investigated using a short DNA substrate comprising a 5'-$^{32}$P radioactively 15mer oligonucleotide annealed onto a 110mer oligonucleotide as explained in Fig. S7.

**Cloning, expression and purification of the *B. subtilis* PykA**

*Cloning of pykA:*

A DNA fragment of 1755 bp carrying the *B. subtilis pykA* gene was amplified from genomic DNA using the PyKAF (5'-TACTTCCAATCCAATGCAAGAAAAACTAAAATTGTTTGTACCATCG-3') and PykAR (5'-TTATCCACTTCCAATGTTATTAAAGAACGCTCGCACG-3') forward and reverse primers, respectively in colony PCR reactions using Q5 high-fidelity DNA polymerase. Typically, a *B. subtilis* single colony was suspended in 20 mL LB and grown at 37$^{o}$C until the optical density reached 0.4-0.8. Thereafter, colony PCR reactions were carried out with genomic DNA (10 μL) acting as the template at 10-fold dilution. The PCR reactions were carried out in a volume of 50 μL with 1 unit Q5 high-fidelity DNA polymerase, 0.5 μM PykAF and PykAR, 0.25 mM dNTPs in 1XQ5 polymerase buffer. PCR products were cleaned up with the Clean-up kit (Thermo Scientific), resolved by agarose electrophoresis, gel extracted using the GeneJET Gel Extraction kit (Thermo Scientific) and cloned into the p2CT plasmid (gift by James Berger) using ligation independent cloning to construct the p2CT-BsuPykA expression vector. This vector codes for an N-terminally His-tagged/MBP PykA protein with the His-MBP tag removable by proteolysis using the TEV protease.

*Expression of PykA:*

For heterologous expression of the *B. subtilis* PykA the p2CT-BsuPykA expression vector was transformed into Rosetta(DE3) *E. coli*. Single colonies were used to inoculate two 600 mL 2xYT cultures, supplemented with 60 µL carbenicillin (50 mg/mL) in 2 Lt conical flasks. The flasks were incubated at 37$^{o}$C, with shaking (180 rpm) until the optical density reached 0.6-0.8. Expression of PykA was induced by the addition of 0.5 mM IPTG and a further 3 hr of growth the cells were harvested by centrifugation at 3,000 g for 15 min. Cells were suspended in 30 mL of buffer A (500 mM NaCl, 50 mM Tris-HCl pH 7.5, 20 mM imidazole) supplemented with 1 mM PMSF and 50 µL protease inhibitor cocktail (Fischer), sonicated at 15 amplitude microns for 1 min, 4 times with 1 min intervals on ice in between. Then benzonase (20 µL) was added to the cell lysate which was further clarified at 40,000 g for 40 min. The soluble crude extract was clarified and filtered through a 0.22 µm filter.

*Purification of PykA:*

The PykA protein was purified from the filtered crude extract using a combination of IMAC (Immobilised Metal Affinity Chromatography) and gel filtration. First, the filtered crude extract was loaded onto a 5 mL HisTrap HP column (GE Healthcare) equilibrated in buffer A. The column was washed thoroughly with buffer A and the PykA protein was eluted using gradient elution with buffer B (500 mM NaCl, 50 mM Tris-HCl pH 7.5, 1 M imidazole). The eluted PykA protein was collected and quantified spectrophotometrically (extinction coefficient 76,780 $M^{-1}$ $cm^{-1}$). TEV protease was added at 1:20 molar ratio while dialyzing the protein solution overnight in dialysis buffer (500 mM NaCl, 50 mM Tris-HCl pH 7.5) at 4$^{o}$C in order to remove the His-MBP tag. The untagged PykA protein was then loaded back onto a 5 mL HisTrap HP column equilibrated in buffer A and the flow-through containing the untagged PykA was collected. Finally, the PykA protein solution was spun concentrated to 5- 7 mL using a vivaspin 10 kDa cutt-off filter. EDTA was added to 1 mM and the PykA was then loaded onto a HiLoad 26/60 Superdex 200 Prep Grade gel filtration column (GE Healthcare) equilibrated in buffer C (500 mM NaCl, 50 mM Tris -HCl pH 7.5 and 1 mM EDTA). Fractions containing the PykA protein were pooled, the PykA was quantified spectrophotometrically (extinction coefficient 8,940 $M^{-1}$ $cm^{-1}$), aliquoted and stored in -80$^{o}$C.

*PykA activity assay:*

The activity of purified PykA was assayed coupling the PykA catalyzed reaction (conversion of phosphoenolpyruvate to pyruvate) to the conversion of pyruvate into lactate catalyzed by LDH (Lactate Dehydrogenase) in the presence of NADH at 25$^{o}$C. The oxidation of NADH to NAD was followed spectrophotometrically at 340 nm and this is directly proportional to the activity of PykA. The LDH-catalyzed reaction was first optimized to ensure that it does not become a limiting factor when measuring the activity of PykA. Then PykA catalyzed reactions were carried out in 96-well plates using the reaction master mix (10 mM Tri-HCl pH 7.5, 10 mM $MgCl_2$, 50 mM KCl, 0.5 mM NADH, 2 mM ADP, 9.375x$10^{-4}$ mg/mL LDH and 5.7 µg/mL PykA, at 25$^{o}$C. Data were analyzed using GraphPad Prism 4 software to plot a Hill plot and a Michaelis-Menten plot from where $V_{max}$, $K_m$ and n (the Hill coefficient) were calculated using GraphPad Prism 4 software.

**Characterization of the oligomeric state of *B. subtilis* PykA**

PykA from all bacteria assembles into a functional tetramer. The oligomeric state of the *B. subtilis* PykA was assessed by native mass spectrometry (MS) and gel filtration. The native MS spectrum revealed a very clean and extremely stable tetramer with miniscule amounts of dimer and monomer (Fig. S3EF). Equally, comparative analytical gel filtration was consistent with a PykA tetramer as it eluted before the γ-globulin (158 kDa) and not as a monomer (62,314 Da) (Fig. S3GH).

**Cloning, expression and purification of the PEPut domain of the *B. subtilis* PykA**

*Cloning of the PEPut domain:*

The DNA fragment coding for the PEPut domain, with the preceding 10 amino acids, was isolated by PCR using genomic DNA and the pepF (5'-TACTTCCAATCCAATGCAGCACAAAATGCAAAAGAAGCT-3') and pepR (5'-TTATCCACTTCCAATGTTATTAAAGAACGCTCGCACG-3') primers, and cloned into the p2CT plasmid, as described above for the *pykA.* The resulting p2CT-PEP expression construct produced an N-terminally His-MBP tagged PEPut protein. The His-MBP tag was removable by proteolysis using the TEV protease.

*Expression and purification of PEPut:*

Expression and purification of the PEPut were carried out as described for the full length PykA protein but the last gel filtration step during purification was omitted. Quantification of the final untagged PEPut (MW 9,582.8 Da) was carried out spectrophotometrically using the extinction coefficients 69,330 $M^{-1}$ $cm^{-1}$ (for the His-MBP tagged PEPut) and 1,490 $M^{-1}$ $cm^{-1}$ (for the untagged PEPut after TEV protease treatment). Purified PEPut is shown in Fig. S6).

**Mass spectrometry**

Native mass spectrometry was carried out with a QToF-1 instrument (Micromass) modified with a 32k quadrupole. The samples were sprayed by nanospray with home-made borosilicate capillary tips back-fitted with platinum wire. The protein was buffer exchanged into 200 mM ammonium acetate using a Zebra desalting spin column (75 µL, 7k MWCO), according to the manufacturer's instructions. The final protein concentration was adjusted to 5 µM and 4 µL of this was loaded per emitter tip. The PykA tetramer peak was isolated in the quadrupole and then activated in the collision cell with 20-160 V against argon gas.

**Acknowledgements**

We thank Nathalie Vega-Czarny and Ioana Popescu for assistance with the MoFlow Astrios cell sorter and the ClarioStar, respectively.

**Conflict of interest**

None declare.

**Funding**

This work was supported by the Biotechnology Biological Sciences Research Council (BBSRC) grant (BB/R013357/1) to P.S. and by the sub-contract (RIS 1165589) to L.J.; M.P. is a PhD student partially funded by a Vice Chancellor's Excellence Award at the University of Nottingham.  S.H. was funded by a PhD fellowship from the MESR (Ministère de l'Enseignement Supérieur et de la Recherche ; ED SDSV, Université Paris-Saclay, Université d'Evry Val d'Essonne). L.J. is on the CNRS (Centre National de la Recherche Scientifique) staff. Funders had no role in the design of the study and in the interpretation of the results.

**Authors' contributions**

L.J. and P.S. designed the study; S.H. and L. J. performed and analyzed *in vivo* experiments; M.P. and A.H. performed *in vitro* experiments; P.S., M.P. and A.H. analyzed *in vitro* data; L.J. and P.S. wrote the paper. All the authors approved the final version of the manuscript.

## FIGURE LEGENDS

**Fig. 1:** Replication defect in *pykA* null cells grown in the MC medium. **A.** Schematic representation of CCM. Glyco: Glycolysis; Gluco: gluconeogenesis; PPP: pentose phosphate pathway; TCA: tricarboxylic acid cycle; O: overflow pathway; G3P: glyceraldehyde 3-phosphate; PEP: phosphoenolpyruvate; Grey arrows: carbon flux; Thick bars: CCM area connected to replication. **B.** Growth rate and pyruvate kinase activity of wild-type and *ΔpykA* cells. Cells were grown exponentially in MC for more than 20 generations using successive dilutions. All along the experiment, growth was assessed by spectrophotometry ($OD_{650nm}$, a typical experiment is shown). Pyruvate kinase activities (mU / mg) in crude extracts were determined from six independent experiments. **C.** *Ori/ter* ratio: *Ori/ter* ratios were determined by qPCR from the genomic DNA of cells collected at $OD_{650nm}$ = 0.10 after exponential growth in MC for more than 10 generations. Numbers in brackets stand for the number of independent measurements (see Material and Methods for details). Mean values and SD are 2.88 +/- 0.07 and 3.03 +/- 0.09 for the WT and *ΔpykA* strain, respectively. The Mann-Whitney U test show that these values are significantly different at $p < .05$ (Table S1). **D.** DNA elongation: Elongation parameters were measured from exponentially growing cells (see above) using a marker frequency analysis determined by qPCR (see Material and Methods for details). The nearly monotonous decrease of marker frequencies from the origin to the terminus showed that there is no pause site along the chromosome (a typical experiment is shown). Numbers in brackets refer to C period (mean and SD from at least 3 experiments; min) and mean fork speed (bp/s). **E.** Number of origins per cell: To determine the number of origins/cell, chloramphenicol was added to cells growing exponentially in MC at $OD_{650nm}$ = 0.15. The drug inhibits replication initiation and cell division but allows completion of ongoing rounds of replication. After 4 hours of drug treatment (runout experiment), cells were analyzed by flow cytometry after DNA staining. Panels: typical runout DNA histograms with the % of cells containing 4 and 8 chromosomes. Numbers in bracket stand for the number of origins/cell (mean and SD from at least 4 experiments). See Material and Methods for details.

**Fig.2:** Replication phenotypes in Cat mutants. **A.** *Ori/ter* ratio analysis. Mutants have been grouped (color code) according to the Mann-Whitney U test at a significance of 0.01 (see Table S1 for details). The horizontal grey bar highlights the wild-type ratio area. Bolded numbers stand for pyruvate kinase activity (mU / mg) in crude extract (mean of three independent measurements, SD/means < 10%). **B.** Other replication parameters. The colored bar is as in panel A. See Fig. 1 and Material and Methods for details.

**Fig. 3:** Replication phenotypes in PEPut and Cat-PEPut interaction mutants. **A.** *Ori/ter* ratio analysis. **B**. Other replication parameters. See Fig. 1-2 and Table S1 for details.

**Fig.4:** *Ori/ter* ratio in *pykA* mutants replicating the genome from *oriN*. Wild-type cells and *pykA* mutants deleted for *oriC* and replicating the genome from *oriN* were grown in MC. The *ori/ter* ratio was measured as in Fig. 1.

**Fig. 5:** PykA stimulates the DNA polymerase activity of DnaE. **A.** Representative alkaline agarose gels showing DnaE primer extension time courses (30, 60, 90, 120 and 150 sec) with DnaE (10 nM) alone and in the presence or absence of PykA (40 nM tetramer) or BSA (40 nM), as described in Materials and Methods. Lanes M and C represent DNA size markers and the control radioactive substrate in the absence of any proteins, respectively. **B.** Representative alkaline agarose gels (from three independent experiments) showing DnaE primer extension time courses (30, 60, 90, 120 and 150 sec) with DnaE (10 nM) and in the presence or absence of PykA (40 nM tetramer), with molar ratios of DnaE monomer:DnaN dimer:HolB monomer:YqeN monomer:DnaX trimer:PykA tetramer set to 1:1:1:1:1:1, considering the oligomeric states of these proteins. **C.** As in B with 2 nM DnaE and molar ratios of 1:1:1:1:1:1.

**Fig. 6:** PykA directly inhibits the helicase activity of the helicase DnaC and via the primase DnaG. **A.** Time courses (5, 10, 15, 25 and 30 min) showing the helicase activity of DnaC/DnaI in the presence or absence of PykA, as indicated. **B.** Time courses (5, 10, 15, 25 and 30 min) showing the helicase activity of DnaC/DnaI in the presence or absence of PykA.and/or DnaG, as indicated. The reactions were carried out as described in Materials and Methods. Representative native PAGE gels (from two independent experiments) are shown with lanes a and b representing control annealed and fully displaced (boiled) control DNA substrates. Data were plotted as a percentage of displaced primer versus time using GrapPad Prism 4 software. Error bars show the standard deviation from two independent repeat experiments.

**Fig. 7:** Model for PykA moonlighting activity in DNA replication: (i) the concentration of signaling metabolites (PEP, ATP, ADP…) is sensed by PykA; (ii) this orchestrates conformational changes in PykA that integrate information originating from Cat, the TSH motif of PEPut (including potential phosphorylation events) and the Cat-PEPut interaction; (iii) PykA then conveys the information to glycolysis and other pathways for regulating CCM, and to the replication receptors DnaE, DnaC and DnaG likely through protein-protein interactions; (iv) changes in receptors' activities modulate initiation and elongation with respect to CCM activity for gating replication in a wide range of growth rates. This signaling model is proposed to be part of the metabolic control of replication.

**FIGURES**

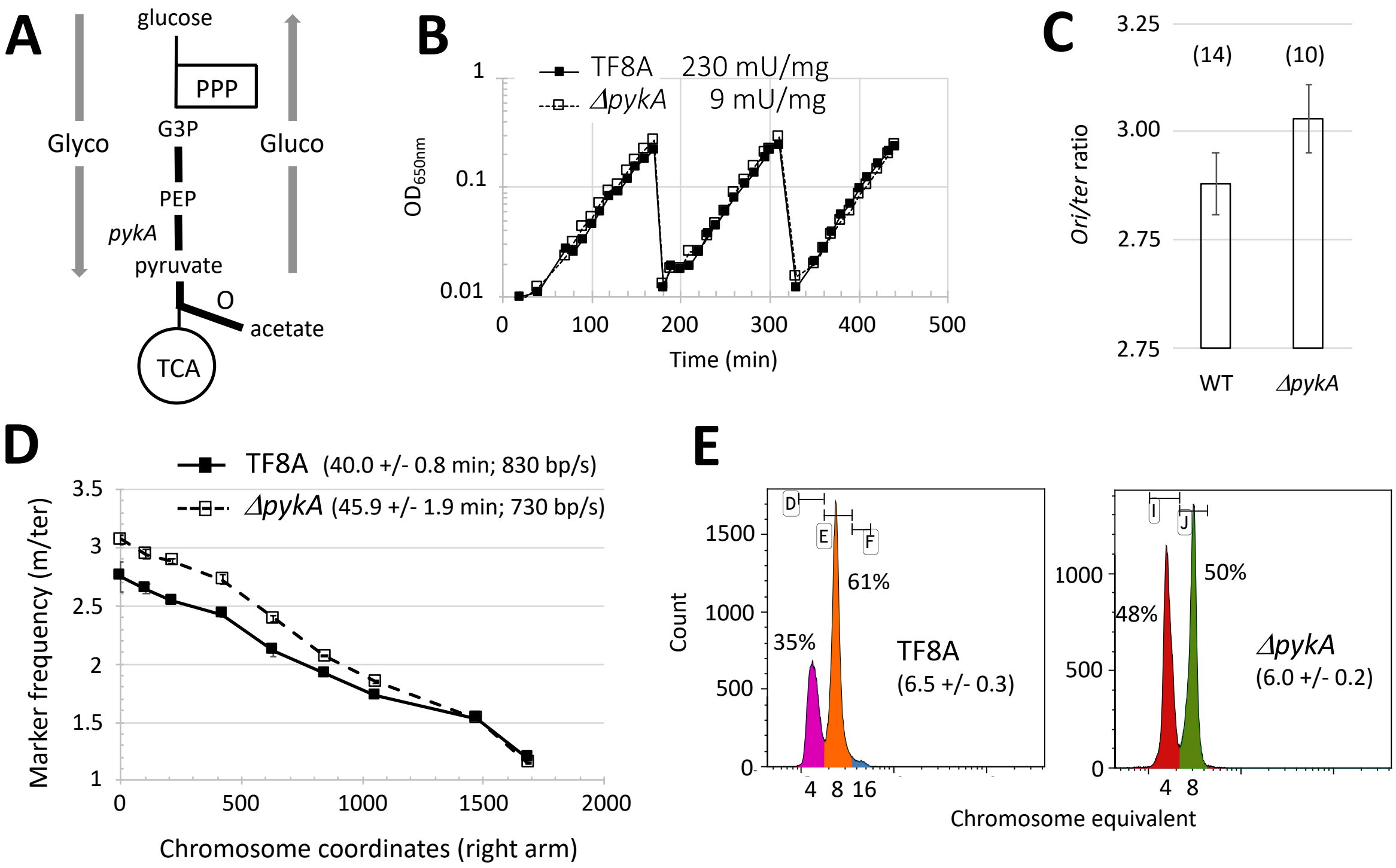

**Fig. 1:** Replication defect in *pykA* null cells.

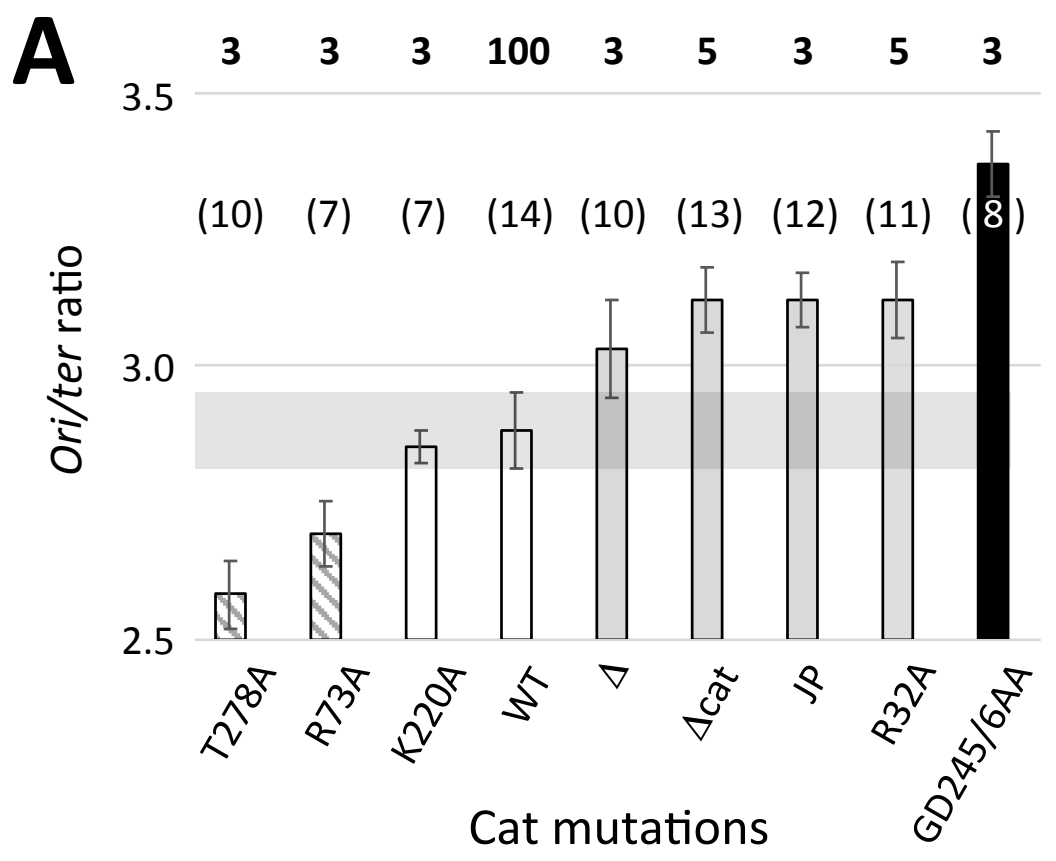

**B**

| Genotype | *Ori/ter* | C period (min) | Fork speed (bp/sec) |
|---|---|---|---|
| $pykA_{T278A}$ | 2.58 +/- 0.06 | 37.3 +/- 0.8 | 870 |
| $pykA_{R73A}$ | 2.69 +/- 0.06 | 41.6 +/- 1.2 | 800 |
| $pykA_{K220A}$ | 2.85 +/- 0.03 | 45.1 +/- 1.0 | 740 |
| WT | 2.88 +/- 0.07 | 40.0 +/- 0.8 | 830 |
| *ΔpykA* | 3.03 +/- 0.09 | 45.9 +/- 1.9 | 730 |
| $pykA_{\Delta cat}$ | 3.12 +/- 0.06 | 55.1 +/- 1.6 | 600 |
| $pykA_{JP}$ | 3.12 +/- 0.05 | 51.2 +/- 0.3 | 650 |
| $pykA_{R32A}$ | 3.12 +/- 0.07 | 49.0 +/- 2.0 | 680 |
| $pykA_{GD246\text{-}6AA}$ | 3.37 +/- 0.06 | 54.5 +/- 1.0 | 610 |

**Fig.2:** Replication phenotypes in Cat mutants.

**A**

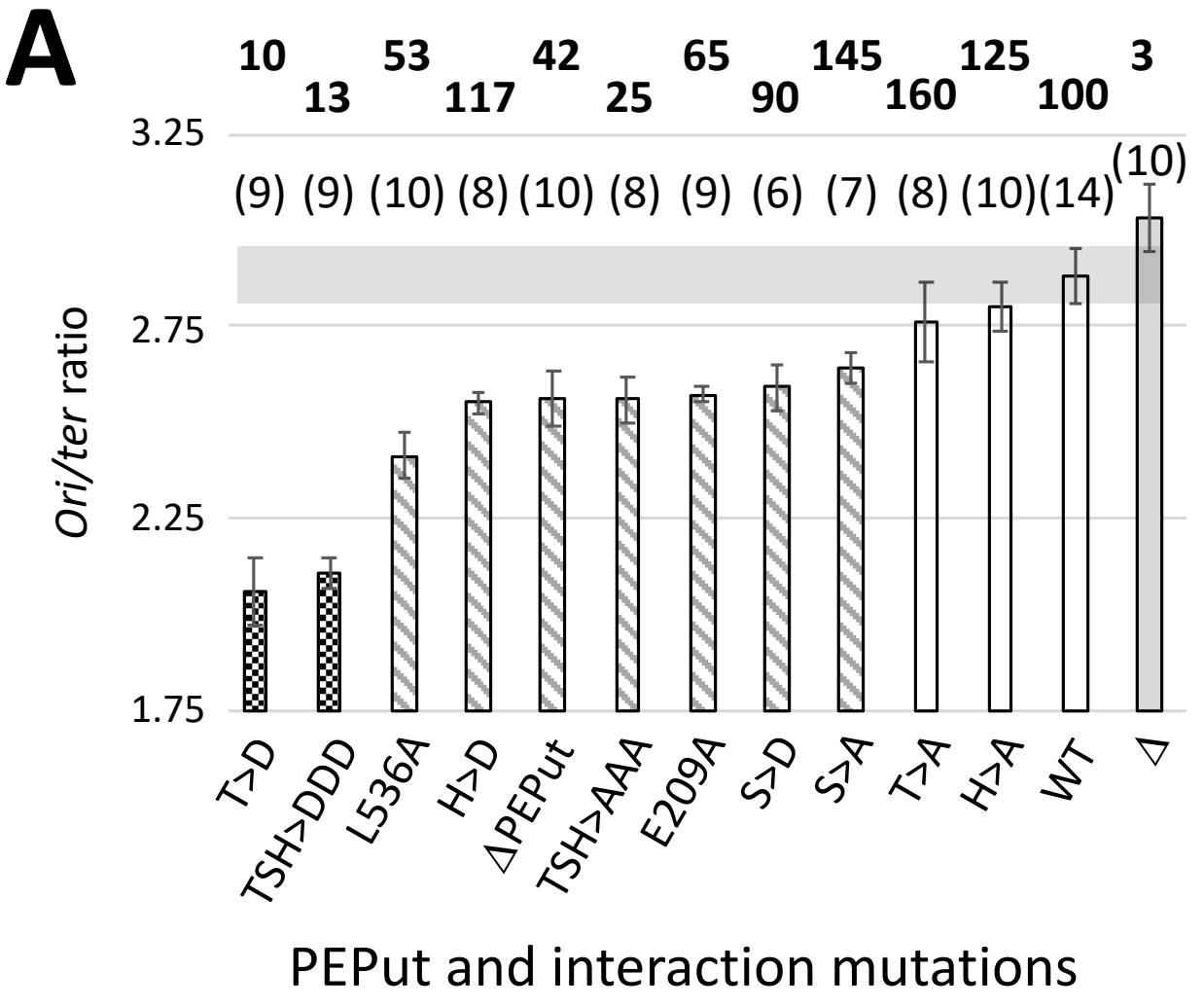

**B**

| Genotype | *Ori/ter* | C period (min) | Fork speed (bp/sec) |
|---|---|---|---|
| *pykA*$_{T>D}$ | 1.97 +/- 0.11 | 31.3 +/- 2.0 | 1060 |
| *pykA*$_{TSH>DDD}$ | 2.08 +/- 0.05 | 33.1 +/- 0.2 | 1010 |
| *pykA*$_{L536A}$ | 2.41 +/- 0.06 | 38.5 +/- 0.5 | 870 |
| *pykA*$_{\Delta PEPut}$ | 2.56 +/- 0.07 | 38.2 +/- 2.3 | 870 |
| *pykA*$_{E209A}$ | 2.57 +/- 0.02 | 42.7 +/- 0.7 | 780 |
| *pykA*$_{TSH>AAA}$ | 2.57 +/- 0.08 | 37.6 +/- 1.2 | 890 |
| *pykA*$_{H>D}$ | 2.59 +/- 0.07 | 39.9 +/- 1.2 | 840 |
| *pykA*$_{S>D}$ | 2.59 +/- 0.05 | 40.9 +/- 1.0 | 810 |
| *pykA*$_{S>A}$ | 2.67 +/- 0.04 | 43.4 +/- 0.9 | 770 |
| *pykA*$_{H>A}$ | 2.79 +/- 0.06 | 44.5 +/- 1.4 | 750 |
| *pykA*$_{T>A}$ | 2.82 +/- 0.12 | 43.3 +/- 0.8 | 770 |
| WT | 2.86 +/- 0.09 | 40.0 +/- 0.8 | 830 |
| *ΔpykA* | 3.09 +/- 0.08 | 45.9 +/- 1.9 | 730 |

**Fig. 3:** Replication phenotypes in PEPut and Cat-PEPut interaction mutants.

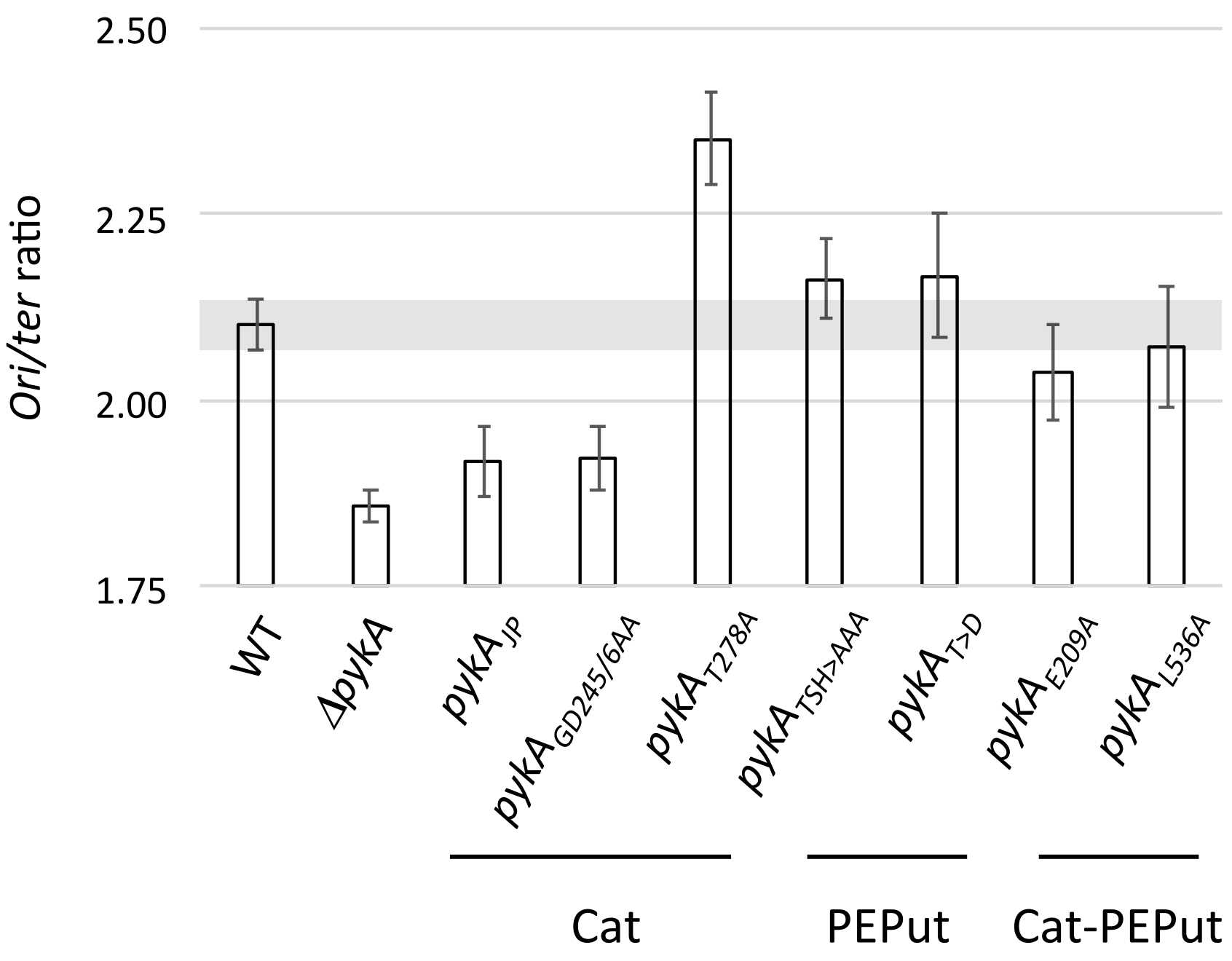

**Fig.4:** *Ori/ter* ratio in *pykA* mutants replicating the genome from *oriN*.

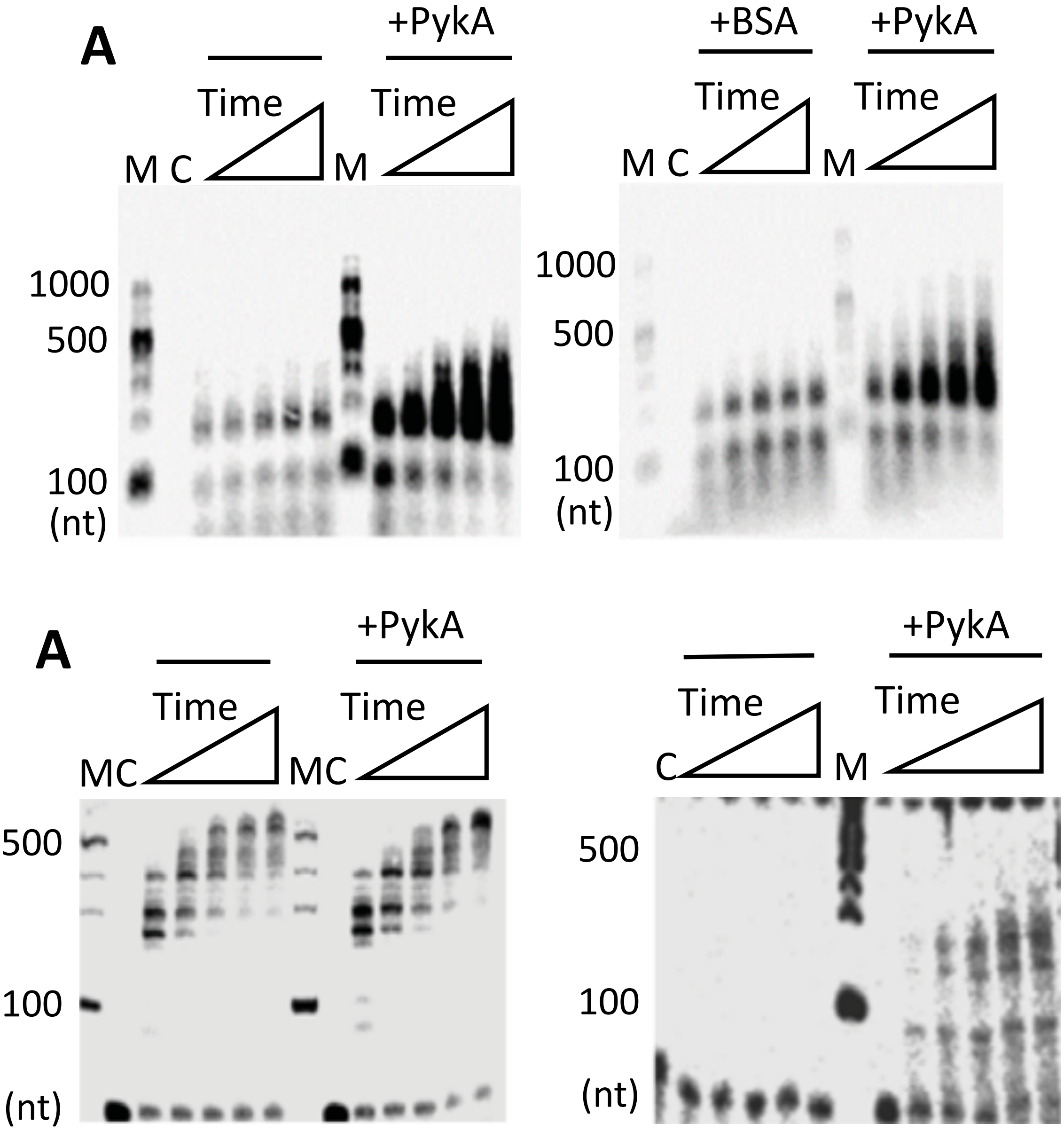

**Fig. 5:** PykA stimulates the DNA polymerase activity of DnaE.

A

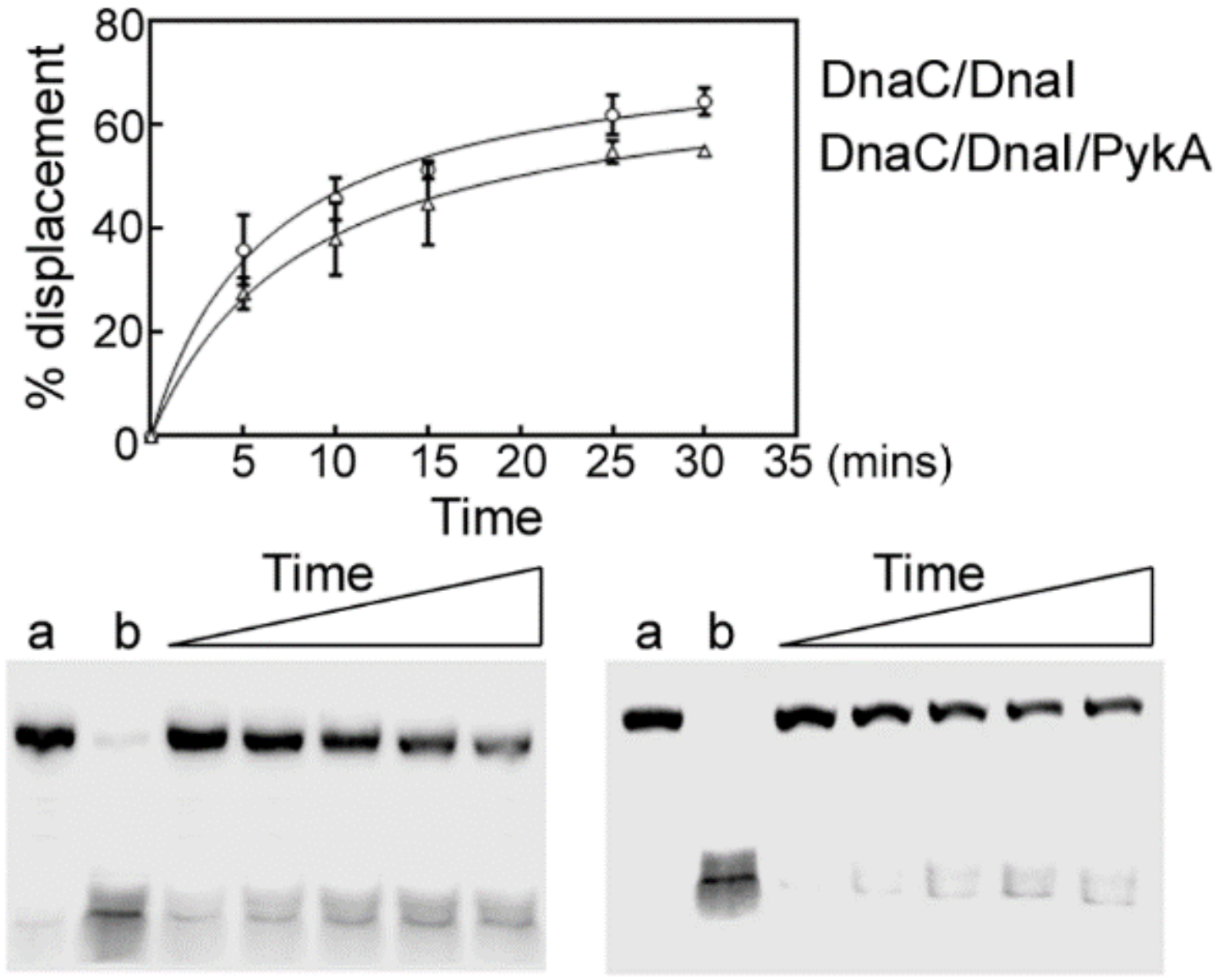

B

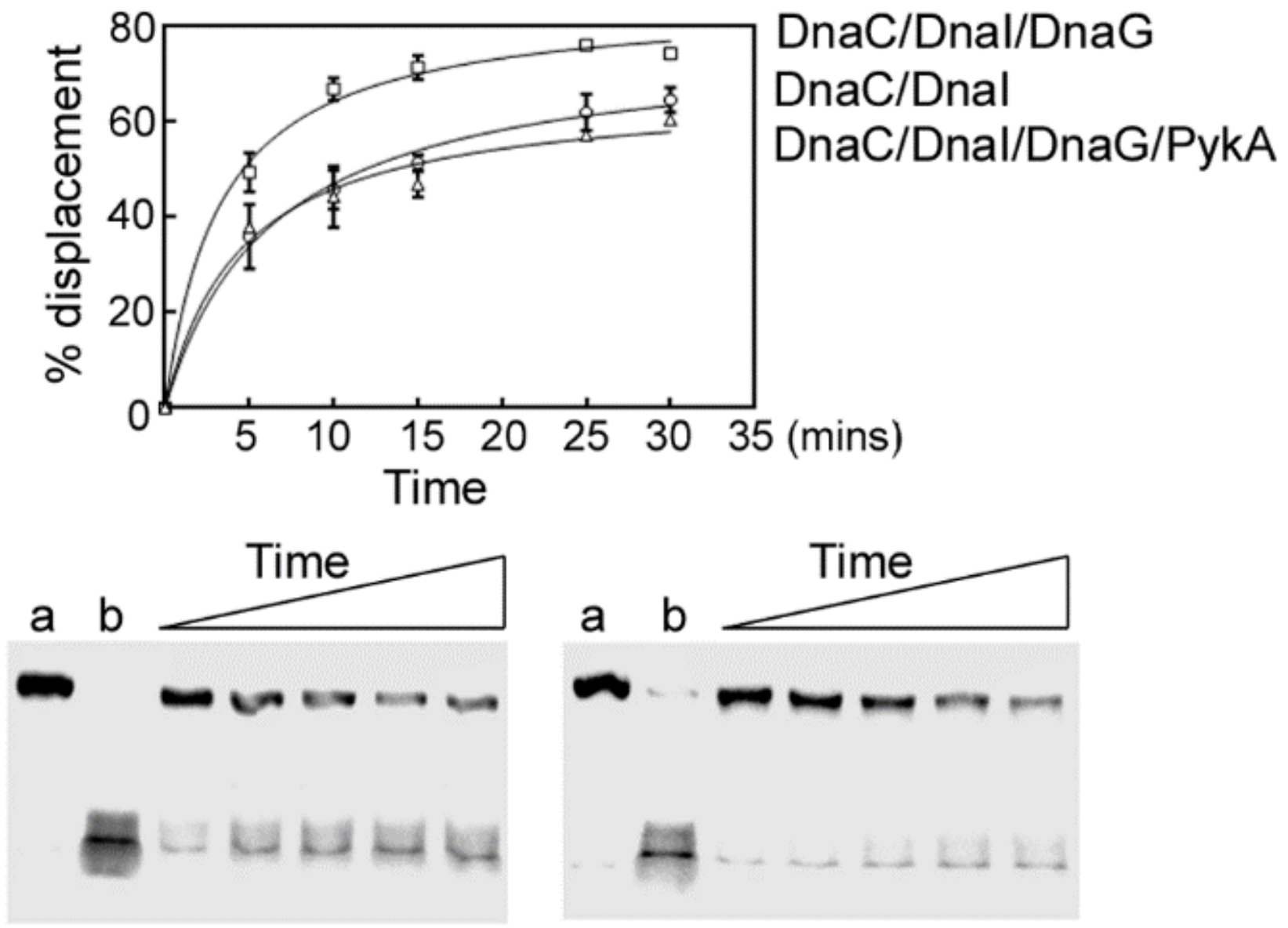

**Fig. 6:** PykA inhibits the helicase activity of the helicase DnaC and cancels the stimulation of DnaC by DnaG.

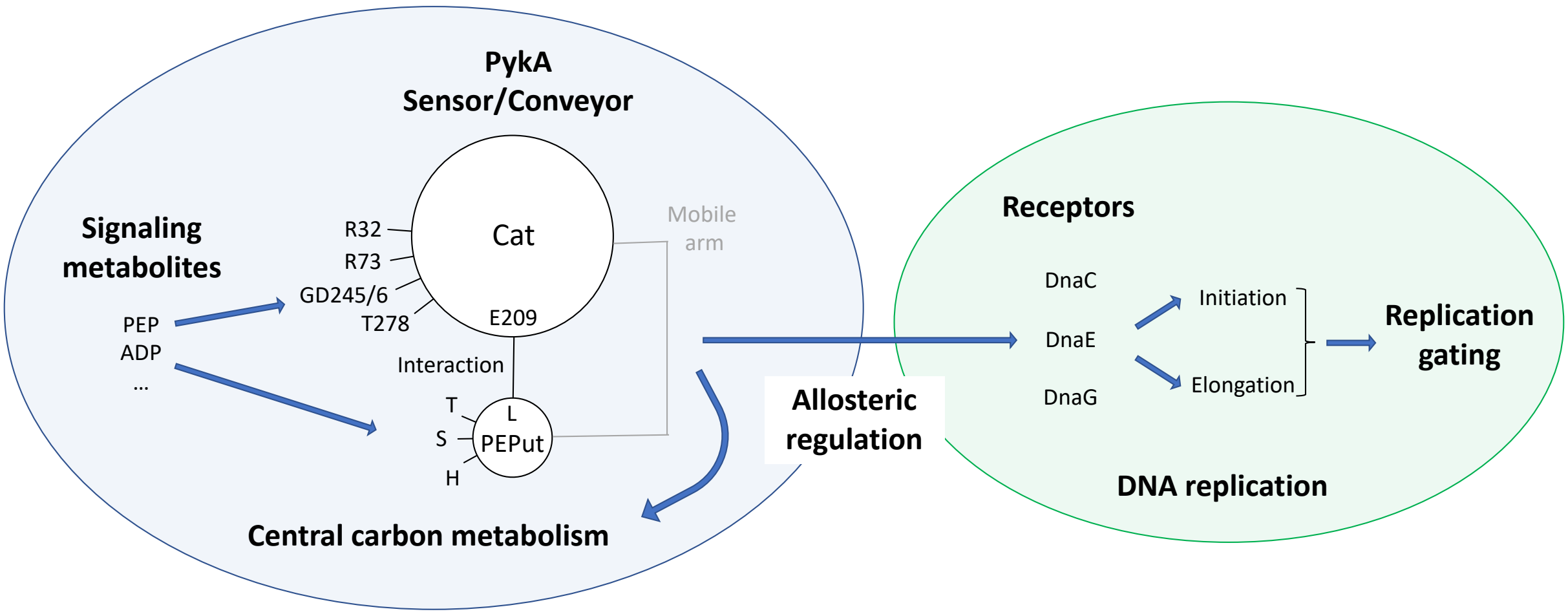

**Fig. 7:** Model for PykA moonlighting activity in DNA replication.

**Table 1:** ori/ter ratios in PykA mutants.

Raw data 1

| | Strain | Main genotype | *Ori/ter* ratio | | |
|---|---|---|---|---|---|
| | | | Data | Mean | SD |
| | TF8A | Wild-type | 3.02<br>2.86<br>3.02<br>3.01<br>2.90<br>2.73<br>2.81<br>2.69<br>3.05<br>2.75<br>2.95<br>2.91<br>2.87<br>2.76 | 2.88 | 0.07 |
| Deletions | DGRM25 | *ΔpykA* | 2.94<br>2.84<br>3.02<br>3.04<br>3.02<br>3.30<br>3.28<br>2.86<br>3.04<br>2.94 | 3.03 | 0.09 |
| | DGRM1017 | *pykA* $_{\Delta cat}$ | 3.15<br>3.04<br>2.98<br>3.12<br>3.22<br>3.34<br>3.19<br>3.19<br>3.04<br>3.18<br>3.09<br>2.98<br>3.06 | 3.12 | 1.00 |
| | DGRM296 | $pykA_{\Delta PEP}$ | 2.71<br>2.62<br>2.60<br>2.41<br>2.55<br>2.47<br>2.39<br>2.69<br>2.71<br>2.47 | 2.56 | 0.07 |

Raw data 2

| | | | | | |
|---|---|---|---|---|---|
| Point mutations in the Cat domain | DGRM24 | *pykA*$_{JP}$ | 3.10<br>3.11<br>3.09<br>3.31<br>3.04<br>2.98<br>3.11<br>3.17<br>3.21<br>3.04<br>3.15<br>3.13 | 3.12 | 0.05 |
| | DGRM1094 | pykA$_{R32A}$ | 3.11<br>2.97<br>3.29<br>3.23<br>3.27<br>3.19<br>2.90<br>3.14<br>3.12<br>2.97<br>3.14 | 3.12 | 0.07 |
| | DGRM1095 | pykA$_{R73A}$ | 2.68<br>2.86<br>2.64<br>2.72<br>2.64<br>2.78<br>2.54 | 2.69 | 0.06 |
| | DGRM1096 | pykA$_{K220A}$ | 2.85<br>2.93<br>2.77<br>2.85<br>2.90<br>2.86<br>2.79 | 2.85 | 0.03 |
| | DGRM1097 | pykA$_{GD245/6AA}$ | 3.46<br>3.44<br>3.52<br>3.33<br>3.34<br>3.40<br>3.21<br>3.24 | 3.37 | 0.06 |
| | DGRM1047 | pykA$_{T278A}$ | 2.63<br>2.36<br>2.69<br>2.68<br>2.65<br>2.57<br>2.63<br>2.53<br>2.54<br>2.51 | 2.58 | 0.06 |

Raw data 3

| | | | | | |
|---|---|---|---|---|---|
| Point mutations in the TSH motif | DGRM302 | pykA$_{TSH>AAA}$ | 2.65 | 2.56 | 0.06 |
| | | | 2.65 | | |
| | | | 2.64 | | |
| | | | 2.44 | | |
| | | | 2.59 | | |
| | | | 2.36 | | |
| | | | 2.62 | | |
| | | | 2.50 | | |
| | DGRM299 | pykA$_{T>A}$ | 3.06 | 2.76 | 0.10 |
| | | | 2.72 | | |
| | | | 2.72 | | |
| | | | 3.00 | | |
| | | | 2.62 | | |
| | | | 2.65 | | |
| | | | 2.62 | | |
| | | | 2.71 | | |
| | DGRM303 | pykA$_{S>A}$ | 2.56 | 2.64 | 0.04 |
| | | | 2.70 | | |
| | | | 2.68 | | |
| | | | 2.72 | | |
| | | | 2.68 | | |
| | | | 2.57 | | |
| | | | 2.58 | | |
| | DGRM298 | pykA$_{H>A}$ | 2.77 | 2.80 | 0.06 |
| | | | 2.63 | | |
| | | | 2.80 | | |
| | | | 2.78 | | |
| | | | 2.96 | | |
| | | | 2.77 | | |
| | | | 2.81 | | |
| | | | 2.98 | | |
| | | | 2.74 | | |
| | | | 2.74 | | |
| | DGRM1019 | pykA$_{TSH>DDD}$ | 1.98 | 2.11 | 0.04 |
| | | | 2.05 | | |
| | | | 2.02 | | |
| | | | 2.17 | | |
| | | | 2.13 | | |
| | | | 2.16 | | |
| | | | 2.13 | | |
| | | | 2.15 | | |
| | | | 2.20 | | |
| | DGRM1018 | pykA$_{T>D}$ | 1.81 | 2.06 | 0.09 |
| | | | 1.93 | | |
| | | | 1.88 | | |
| | | | 2.20 | | |
| | | | 2.10 | | |
| | | | 2.28 | | |
| | | | 2.12 | | |
| | | | 2.07 | | |
| | | | 2.19 | | |
| | DGRM1016 | pykA$_{S>D}$ | 2.50 | 2.59 | 0.06 |
| | | | 2.60 | | |
| | | | 2.66 | | |
| | | | 2.44 | | |
| | | | 2.72 | | |
| | | | 2.60 | | |
| | DGRM1015 | pykA$_{H>D}$ | 2.50 | 2.55 | 0.07 |
| | | | 2.52 | | |
| | | | 2.61 | | |
| | | | 2.78 | | |
| | | | 2.52 | | |
| | | | 2.56 | | |
| | | | 2.51 | | |
| | | | 2.38 | | |

Raw data 4

| | | | | | |
|---|---|---|---|---|---|
| Cat-PEPut interaction mutants | DGRM1046 | $pykA_{E209A}$ | 2.56<br>2.56<br>2.54<br>2.64<br>2.57<br>2.56<br>2.59<br>2.61<br>2.54 | 2.57 | 0.02 |
| | DGRM1048 | $pykA_{L536A}$ | 2.36<br>2.45<br>2.43<br>2.36<br>2.36<br>2.64<br>2.35<br>2.30<br>2.47<br>2.35 | 2.41 | 0.06 |

**Table 1**

Mann-Whitney U test analysis

Ori/ter ratios were compared used the Mann-Whitney U test (https://www.socscistatistics.com/test/mannWhitney.defaut2.aspx) with a significance level of 0.01 and the two-tailed hypothesis.

**Data analysis versus the wild-type strain**

| | | TF8A | *ΔpykA* [a,b] | $pykA_{\Delta Cat}$ | $pykA_{JP}$ | $pykA_{R32A}$ | $pykA_{R73A}$ | $pykA_{K220A}$ | $pykA_{GD245/6AA}$ | $pykA_{T278A}$ |
|---|---|---|---|---|---|---|---|---|---|---|
| TF8A | Data *U*-value | 98 | 33.5 | 2 | 6 | 14.5 | 10.5 | 40 | 0 | NR |
| | Critical *U*-value | 42 | 36 | 15 | 34 | 30 | 15 | 15 | 18 | NR |
| | *p*-value | 0.9844 | 0.03486 | 0.00052 | 0.00006 | 0.0007 | 0.00466 | 0.5287 | 0.00016 | < 0.00001 |

| | | $pykA_{\Delta PEP}$ | $pykA_{TSH>AAA}$ | $pykA_{T>A}$ | $pykA_{S>A}$ | $pykA_{H>A}$ | $pykA_{TSH>DDD}$ | $pykA_{T>D}$ | $pykA_{S>D}$ | $pykA_{H>D}$ | $pykA_{K209A}$ | $pykA_{L536A}$ |
|---|---|---|---|---|---|---|---|---|---|---|---|---|
| TF8A | Data *U*-value | 2.5 | 0 | 27 | 2 | 45 | 0 | 0 | 1 | 4 | 0 | NR |
| | Critical *U*-value | 26 | 18 | 18 | 15 | 26 | 22 | 22 | 11 | 18 | 22 | NR |
| | *p*-value | 0.00008 | 0.00016 | 0.05118 | 0.00052 | 0.15272 | 0.00008 | 0.00008 | 0.00084 | 0.00044 | 0.00008 | < 0.00001 |

[a]: significance level of 0.05 (not significant at a significance level of 0.01)

[b]: pairs of data significantly different are highlighted in grey

**Pairwise analysis of Cat mutants with an altered ratio**

| | *ΔpykA* | $pykA_{\Delta Cat}$ | $pykA_{JP}$ | $pykA_{R32A}$ | $pykA_{R73A}$ | $pykA_{GD245/6AA}$ | $pykA_{T278A}$ |
|---|---|---|---|---|---|---|---|
| *ΔpykA* | 50 (16) [c] 0.9681 | 35 (24) 0.06724 | 28 (21) 0.03752 | 35 (18) 0.17068 | 1.5 (9) 0.00128 | 4 (11) 0.00164 | 0 (16) 0.00018 |
| $pykA_{\Delta Cat}$ | | 84.5 (34) 0.97606 | 77 (31) 0.97606 | 70.5 (27) 0.97606 | 0 (13) 0.00036 | 4.5 (17) 0.00068 | NR < 0.00001 |
| $pykA_{JP}$ | | | 72 (27) 0.97606 | 60.5 (24) 0.75656 | 0 (12) 0.00044 | 2.5 (15) 0.00052 | 0 (21) 0.00008 |
| $pykA_{R32A}$ | | | | 60.5 (21) 0.97606 | 0 (10) 0.00058 | 5 (13) 0.00148 | 0 (18) 0.00012 |
| $pykA_{R73A}$ | | | | | 24.5 (4) 0.95216 | 0 (6) 0.00148 | 14 (9) 0.0455 |
| $pykA_{GD245/6AA}$ | | | | | | 32 (7) 0.96012 | 0 (11) 0.00044 |
| $pykA_{T278A}$ | | | | | | | 50 (16) 0.9681 |

[c]: Data *U*-value; (critical *U*-value); *p*-value

**Pairwise analysis of PEPut mutants with an altered ratio**

| | $pykA_{\Delta PEP}$ | $pykA_{TSH>AAA}$ | $pykA_{S>A}$ | $pykA_{TSH>DDD}$ | $pykA_{T>D}$ | $pykA_{S>D}$ | $pykA_{H>D}$ | $pykA_{K209A}$ | $pykA_{L536A}$ | pykAR73A | pykAT278A |
|---|---|---|---|---|---|---|---|---|---|---|---|
| pykADPEP | 50 (16) 0.9681 | 38 (11) 0.89656 | 23 (9) 0.26272 | 0 (13) 0.00028 | 0 (13) 0.00028 | 27 (6) 0.78716 | 37 (11) 0.82588 | 44 (13) 0.9681 | 14 (16) 0.0736 | 15 (9) 0.05744 | 47.5 (16) 0.88076 |
| pykATSH>AAA | | 32 (7) 0.96012 | 15 (6) 0.14706 | 0 (9) 0.00062 | 0 (9) 0.00062 | 20.5 (4) 0.69654 | 27.5 (7) 0.67448 | 31 (9) 0.6672 | 12.5 (11) 0.0164 | 9 (6) 0.03236 | 35.5 (11) 0.71884 |
| pykAS>A | | | 24.5 (4) 0.95216 | 0 (7) 0.00104 | 0 (7) 0.00104 | 15.5 (3) 0.47777 | 10.5 (6) 0.04884 | 12 (7) 0.04444 | 3 (9) 0.00214 | 18.5 (4) 0.48392 | 19.5 (9) 0.1443 |
| pykATSH>DDD | | | | 40.5 (11) 0.9681 | 34.5 (11) 0.62414 | 0 (5) 0.0018 | 0 (9) 0.00062 | 0 (11) 0.00042 | 0 (13) 0.00028 | 0 (7) 0.0014 | 0 (13) 0.00028 |
| pykAT>D | | | | | 40.5 (11) 0.9681 | 0 (5) 0.0018 | 0 (9) 0.00062 | 0 (11) 0.00042 | 0 (13) 0.00028 | 0 (7) 0.0014 | 0 (13) 0.00028 |
| pykAS>D | | | | | | 18 (2) 0.93624 | 19.5 (4) 0.60306 | 22 (5) 0.59612 | 6 (6) 0.01078 | 9.5 (3) 0.11642 | 30 (6) 0.96012 |
| pykAH>D | | | | | | | 32 (7) 0.96012 | 20.5 (9) 0.14986 | 10 (11) 0.0088 | 7 (6) 0.01778 | 26.5 (11) 0.24604 |
| pykAK209A | | | | | | | | 40.5 (11) 0.9681 | 8.5 (13) 0.00328 | 9 (7) 0.0198 | 40.5 (13) 0.7414 |
| pykAL536A | | | | | | | | | 50 (16) 0.9681 | 2 (9) 0.00152 | 11.5 (16) 0.0041 |
| pykAR73A | | | | | | | | | | 24.5 (4) 0.95216 | 14 (9) 0.0455 |
| PykAT278A | | | | | | | | | | | 50 (216) 0.9681 |

# Table 2

## Strains

| Species | Types of strains | Genotype | Main phenotypes | Name | Context | Source of construction | Obtained by transformation with: |
|---|---|---|---|---|---|---|---|
| *B. subtilis* | Parental strains | *trpC2* | wild-type | TF8A | | Jannière et al, 2007 | |
| | | *pykA-tet* | TetR marker just downstream of *pykA* | DGRM295 | TF8A | This work | PCR product ➔ TF8A (Tet)[1] |
| | | *pykAFLAG-tet* | PykA C-terminally fused the the FLAG-tag, TetR | DGRM1098 | TF8A | This work | PCR product ➔ TF8A (Tet) |
| | | *pykA::spc trpC2* | Deletion of *pykA*, SpR | DGRM25 | TF8A | Jannière et al, 2007 | |
| | | *spoIIIJ::oriN-cat DoriC* | OriN$^{+}$ deleted for *oriC* CmR | DGRM589 | TF8A | Nouri et al., 2018 | |
| | Cat mutants | *pykA*$_{Dcat}$*-tet trpC2* | Encodes the PEPut domain of PykA (Cat deleted), TetR | DGRM1017 | TF8A | This work | PCR product ➔ TF8A (Tet) |
| | | *pykA*$_{R32A}$*-tet trpC2* | Mutation R32A in the Cat domain, TetR | DGRM1094 | TF8A | This work | PCR product ➔ TF8A (Tet) |
| | | *pykA*$_{R73A}$*-tet trpC2* | Mutation R37A in the Cat domain, TetR | DGRM1095 | TF8A | This work | PCR product ➔ TF8A (Tet) |
| | | *pykA*$_{K220A}$*-tet trpC2* | Mutation K220A in the Cat domain, TetR | DGRM1096 | TF8A | This work | PCR product ➔ TF8A (Tet) |
| | | *pykA*$_{GD245\text{-}246AA}$*-tet trpC2* | Mutation GD245-246AA in the Cat domain, TetR | DGRM1097 | TF8A | This work | PCR product ➔ TF8A (Tet) |
| | | *pykA*$_{T278A}$*-tet trpC2* | Mutation T278A in the Cat domain, TetR | DGRM1047 | TF8A | This work | PCR product ➔ TF8A (Tet) |
| | | *pykAJP-prm trpC2* | Deletion of 27 amino acids (208-234) in the Cat domain, PhlR | DGRM24 | TF8A | Jannière et al, 2007 | |
| | PEPut mutants | *pykA*$_{DPEP}$*-tet trpC2* | Encodes the catalytic domain of PykA (PEPut deleted), TetR | DGRM296N | TF8A | This work | PCR product ➔ TF8A (Tet) |
| | | *pykA*$_{T>A}$*-tet trpC2* | Mutation T537A in the PEPut domain, TetR | DGRM299 | TF8A | This work | PCR product ➔ TF8A (Tet) |
| | | *pykA*$_{S>A}$*-tet trpC2* | Mutation S538A in the PEPut domain, TetR | DGRM303 | TF8A | This work | PCR product ➔ TF8A (Tet) |
| | | *pykA*$_{H>A}$*-tet trpC2* | Mutation H539A in the PEPut domain, TetR | DGRM298 | TF8A | This work | PCR product ➔ TF8A (Tet) |
| | | *pykA*$_{TSH>AAA}$*-tet trpC2* | Mutation TSH537-539AAA in the PEPut domain, TetR | DGRM302 | TF8A | This work | PCR product ➔ TF8A (Tet) |
| | | *pykA*$_{T>D}$*-tet trpC2* | Mutation T537D in the PEPut domain, TetR | DGRM1018 | TF8A | This work | PCR product ➔ TF8A (Tet) |
| | | *pykA*$_{S>D}$*-tet trpC2* | Mutation S538D in the PEPut domain, TetR | DGRM1016 | TF8A | This work | PCR product ➔ TF8A (Tet) |
| | | *pykA*$_{H>D}$*-tet trpC2* | Mutation H539D in the PEPut domain, TetR | DGRM1015 | TF8A | This work | PCR product ➔ TF8A (Tet) |
| | | *pykA*$_{TSH>DDD}$*-tet trpC2* | Mutation TSH537-539DDD in the PEPut domain, TetR | DGRM1019 | TF8A | This work | PCR product ➔ TF8A (Tet) |
| | Cat-PEPut interaction mutants | *pykAE209*$_{A}$*-tet trpC2* | Mutation E209A in the Cat domain, TetR | DGRM1046 | TF8A | This work | PCR product ➔ TF8A (Tet) |
| | | *pykA*$_{L536A}$*-tet trpC2* | Mutation L536A in the PEPut domain, TetR | DGRM1048 | TF8A | This work | PCR product ➔ TF8A (Tet) |
| | *oriN* mutants | *spoIIIJ::oriN-cat DoriC pykA::spc* | OriN$^{+}$ deleted for *oriC* and *pykA* CmR SpR | DGRM1140 | TF8A | This work | DGRM25 ➔ DGRM589 (Tet) |
| | | *spoIIIJ::oriN-cat DoriC pykAJP-prm* | OriN$^{+}$ deleted for *oriC* and residues 208-234 of the Cat domain, CmR PhlR | DGRM1141 | TF8A | This work | DGRM24 ➔ DGRM589 (Tet) |
| | | *spoIIIJ::oriN-cat DoriC pykA*$_{GD245/6AA}$*-tet* | OriN$^{+}$ deleted for *oriC* and pykA$_{GD245/6AA}$ mutation in the Cat domain, CmR TetR | DGRM1142 | TF8A | This work | DGRM1097 ➔ DGRM589 (Tet) |
| | | *spoIIIJ::oriN-cat DoriC pykA*$_{T278A}$*-tet* | OriN$^{+}$ deleted for *oriC* and pykA$_{T278A}$ mutation in the Cat domain, CmR TetR | DGRM1146 | TF8A | This work | DGRM1047 ➔ DGRM589 (Tet) |
| | | *spoIIIJ::oriN-cat DoriC pykA*$_{TSH>AAA}$*-tet* | OriN$^{+}$ deleted for *oriC* and pykA$_{TSH>AAA}$ mutation in the PEPut domain, CmR TetR | DGRM1143 | TF8A | This work | DGRM302 ➔ DGRM589 (Tet) |
| | | *spoIIIJ::oriN-cat DoriC pykA*$_{T>D}$*-tet* | OriN$^{+}$ deleted for *oriC* and pykA$_{T>D}$ mutation in the PEPut domain, CmR TetR | DGRM1144 | TF8A | This work | DGRM10148 ➔ DGRM589 (Tet) |
| | | *spoIIIJ::oriN-cat DoriC pykA*$_{E209A}$*-tet* | OriN$^{+}$ deleted for *oriC* and pykA$_{E209A}$ mutation in the Cat domain, CmR TetR | DGRM1145 | TF8A | This work | DGRM1046 ➔ DGRM589 (Tet) |
| | | *spoIIIJ::oriN-cat DoriC pykA*$_{L536A}$*-tet* | OriN$^{+}$ deleted for *oriC* and pykA$_{L536A}$ mutation in the PEPut domain, CmR TetR | DGRM1147 | TF8A | This work | DGRM1048 ➔ DGRM589 (Tet) |

[1]: X ➔ Y indicates that strain Y was transformed with DNA from source X using the selection indicated in brakets.

**REFERENCES**

Alpert, C.A., Frank, R., Stüber, K., Deutscher, J., and Hengstenberg, W. (1985). Phosphoenolpyruvate-dependent protein kinase enzyme I of *Streptococcus faecalis*: purification and properties of the enzyme and characterization of its active center. Biochemistry *24*, 959–964.

Baranska, S., Glinkowska, M., Herman-Antosiewicz, A., Maciag-Dorszynska, M., Nowicki, D., Szalewska-Palasz, A., Wegrzyn, A., and Wegrzyn, G. (2013). Replicating DNA by cell factories: roles of central carbon metabolism and transcription in the control of DNA replication in microbes, and implications for understanding this process in human cells. Microb. Cell Fact. *12*, 55.

Bipatnath, M., Dennis, P.P., and Bremer, H. (1998). Initiation and velocity of chromosome replication in *Escherichia coli* B/r and K-12. J. Bacteriol. *180*, 2675–2273.

Bollenbach, T.J., Mesecar, A.D., and Nowak, T. (1999). Role of lysine 240 in the mechanism of yeast pyruvate kinase catalysis. Biochemistry *38*, 9137–9145.

Boukouris, A.E., Zervopoulos, S.D., and Michelakis, E.D. (2016). Metabolic enzymes moonlighting in the nucleus: metabolic regulation of gene transcription. Trends Biochem. Sci. *41*, 712–730.

Boye, E., and Nordström, K. (2003). Coupling the cell cycle to cell growth. EMBO Reports *4*, 757–760.

Bruck, I., and O'Donnell, M. (2000). The DNA replication machine of a gram-positive organism. J. Biol. Chem. *275*, 28971–28983.

Buchakjian, M.R., and Kornbluth, S. (2010). The engine driving the ship: metabolic steering of cell proliferation and death. Nat. Rev. Microbiol. *11*, 715–727.

Burnell, J.N., and Hatch, M.D. (1984). Regulation of $C_4$ photosynthesis: Identification of a catalycally important histidine residue and its role in the regulation of pyruvate,$P_i$ dikinase. Arch. Biochem. Biophys. *231*, 175–182.

Burnell, J.N. (2010). Cloning and characterization of Escherichia coli DUF299: a bifunctional ADP-dependent kinase - $P_i$-dependent pyrophosphorylase from bacteria. BMC Biochem. *11*, 1–8.

Burnell, J.N., and Chastain, C.J. (2006). Cloning and expression of maize-leaf pyruvate, Pi dikinase regulatory protein gene. Biochem. Biophys. Res. Commun. *345*, 675–680.

Burnetti, A.J., Aydin, M., and Buchler, N.E. (2015). Cell cycle Start is coupled to entry into the yeast metabolic cycle across diverse strains and growth rates. Mol. Biol. Cell *27*, 64–74.

Cai, L., Sutter, B.M., Li, B., and Tu, B.P. (2011). Acetyl-CoA induces cell growth and proliferation by promoting the acetylation of histones at growth genes. Mol. Cell *42*, 426–437.

Chuang, C., Prasanth, K.R., and Nagy, P.D. (2017). The glycolytic pyruvate kinase is recruited directly into the viral replicase complex to generate ATP for RNA synthesis. Cell Host Microbe *22*, 639–652.e7.

Dai, R.P., Yu, F.X., Goh, S.R., Chng, H.W., Tan, Y.L., Fu, J.L., Zheng, L., and Luo, Y. (2008). Histone 2B (H2B) expression is confined to a proper NAD+/NADH redox status. J. Biol. Chem. *283*, 26894–26901.

Dervyn, E., Suski, C., Daniel, R., Bruand, C., Chapuis, J., Errington, J., Jannière, L., and Ehrlich, D.S. (2001). Two essential DNA polymerases at the bacterial replication fork. Science *294*, 1716–1719.

Dickinson, J.R., and Williams, A.S. (1987). The *cdc30* mutation in *Saccharomyces cerevisiae* results in a temperature-sensitive isoenzyme of phosphoglucose isomerase. J. Gen. Microbiol. *133*, 135–140.

Du, Y.-C.N., and Stillman, B. (2002). Yph1p, an ORC-Interacting Protein: Potential Links between Cell Proliferation Control, DNA Replication, and Ribosome Biogenesis. Cell *109*, 835–848.

Ewald, J.C. (2018). How yeast coordinates metabolism, growth and division. Curr. Opin. Microbiol. *45*, 1–7.

Eymann, C., Dreisbach, A., Albrecht, D., Bernhardt, J., Becher, D., Gentner, S., Tam, L.T., Büttner, K., Buurman, G., Scharf, C., et al. (2004). A comprehensive proteome map of growing *Bacillus subtilis* cells. Proteomics *4*, 2849–2876.

Flåtten, I., Fossum-Raunehaug, S., Taipale, R., Martinsen, S., and Skarstad, K. (2015). The DnaA protein is not the limiting factor for initiation of replication in *Escherichia coli*. PLoS Genet. *11*, e1005276.

Fornalewicz, K., Wieczorek, A., Wegrzyn, G., and Łyżeń, R. (2017). Silencing of the pentose phosphate pathway genes influences DNA replication in human fibroblasts. Gene *635*, 33–38.

Goss, N.H., Evans, C.T., and Wood, H.G. (1980). Pyruvate phosphate dikinase: sequence of the histidyl peptide, the pyrophosphoryl and phosphoryl carrier. Biochemistry *19*, 5805–5809.

Grosse, F., Nasheuer, H.-P., Schltissek, S., and Schomburg, U. (1986). Lactate dehydrogenase and glyceraldehyde-phosphate dehydrogenase are single-stranded DNA-binding proteins that affect the DNA-polymerase-α–primase complex. Eur. J. Biochem. *160*, 459–467.

Hassan, A.K.M., Moriya, S., Ogura, M., Tanaka, T., Kawamura, F., and Ogasawara, N. (1997). Suppression of initiation defects of chromosome replication in *Bacillus subtilis dnaA* and *oriC*-deleted mutants by integration of a plasmid replicon into the chromosomes. J. Bacteriol. *179*, 2494–2502.

Helmstetter, C.E. (1996). *Timing of synthetic activities in the cell cycle*. In *Escherichia Coli* And *Salmonella*, F.C. Neidhart, R.I. Curtis, E.C. Ingraham, K.B. Lin, and L.E.C.C., eds. (Washington DC: ASM Press), pp. 1591–1605.

Hernandez, J.V., and Bremer, H. (1993). Characterization of RNA and DNA synthesis in *Escherichia coli* strains devoid of ppGpp. J. Biol. Chem. *268*, 10851–10862.

Herzberg, O., Chen, C.C., Kapadia, G., McGuire, M., Carroll, L.J., Noh, S.J., and Dunaway-Mariano, D. (1996). Swiveling-domain mechanism for enzymatic phosphotransfer between remote reaction sites. Proc. Natl. Aca. Sci. USA *93*, 2652–2657.

Hu, C.-M., Tien, S.-C., Hsieh, P.-K., Jeng, Y.-M., Chang, M.-C., Chang, Y.-T., Chen, Y.-J., Chen, Y.-J., Lee, E.Y.H.P., and Lee, W.-H. (2019). High Glucose Triggers Nucleotide Imbalance through O-GlcNAcylation of Key Enzymes and Induces KRAS Mutation in Pancreatic Cells. Cell Metab. *29*, 1334–1349.e10.

Hughes, P., Landoulsi, A., and Kohiyama, M. (1988). A novel role for cAMP in the control of the activity of the *E. coli* chromosome replication initiator protein, DnaA. Cell *55*, 343–350.

Ishida, T., Akimitsu, N., Kashioka, T., Hatano, M., Kubota, T., Ogata, Y., Sekimizu, K., and Katayama, T. (2004). DiaA, a novel DnaA-binding protein, ensures the timely initiation of *Escherichia coli* chromosome replication. J. Biol. Chem. *279*, 45546–45555.

Jannière, L., Canceill, D., Suski, C., Kanga, S., Dalmais, B., Lestini, R., Monnier, A.-F., Chapuis, J., Bolotin, A., Titok, M., et al. (2007). Genetic evidence for a link between glycolysis and DNA replication. PLoS ONE *2*, e447.

Jindal, H.K., and Vishwanatha, J.K. (1990). Functional identity of a primer recognition protein as phosphoglycerate kinase. J. Biol. Chem. *265*, 6540–6543.

Kim, J.-W., and Dang, C.V. (2005). Multifaceted roles of glycolytic enzymes. Trends Biochem. Sci. *30*, 142–150.

Klevecz, R.R., Bolen, J., Forrest, G., and Murray, D.B. (2004). A genomewide oscillation in transcription gates DNA replication and cell cycle. Proc. Natl. Aca. Sci. USA *101*, 1200–1205.

Konieczna, A., Szczepańska, A., Sawiuk, K., Wegrzyn, G., and Łyżeń, R. (2015a). Effects of partial silencing of genes coding for enzymes involved in glycolysis and tricarboxylic acid cycle on the enterance of human fibroblasts to the S phase. BMC Cell Biol. *16*, 16.

Konieczna, A., Szczepańska, A., Sawiuk, K., Łyżeń, R., and Wegrzyn, G. (2015b). Enzymes of the central carbon metabolism: Are they linkers between transcription, DNA replication, and carcinogenesis? Med. Hypotheses *84*, 58–67.

Krause, K., Maciag-Dorszynska, M., Wosinski, A., Gaffke, L., Morcinek-Orłowska, J., Rintz, E., Bielańska, P., Szalewska-Pałasz, A., Muskhelishvili, G., and Wegrzyn, G. (2020). The role of metabolites in the link between DNA replication and central carbon metabolism in *Escherichia coli*. Genes *11*, 447.

Laffan, J., and Firshein, W. (1987). Membrane protein binding to the origin region of *Bacillus subtilis*. J. Bacteriol. *169*, 4135–4140.

Laffan, J., and Firshein, W. (1988). Origin-specific DNA-binding membrane-associated protein may be involved in repression of initiation in *Bacillus subtilis*. Proc. Natl. Aca. Sci. USA *85*, 7452–7456.

Le Chatelier, E., Becherel, O.J., d'Alencon, E., Canceill, D., Ehrlich, S.D., Fuchs, R.P.P., and Janniere, L. (2004). Involvement of DnaE, the second replicative DNA polymerase from *Bacillus subtilis*, in DNA mutagenesis. J. Biol. Chem. *279*, 1757–1767.

Li, X., Qian, X., Jiang, H., Xia, Y., Zheng, Y., Li, J., Huang, B.-J., Fang, J., Qian, C.-N., Jiang, T., et al. (2018). Nuclear PGK1 alleviates ADP-dependent inhibition of CDC7 to promote DNA replication. Mol. Cell *72*, 650–660.e658.

Lu, M., Campbell, J.L., Boye, E., and Kleckner, N. (1994). SeqA: A negative modulator of replication nitiation in *E. coli*. Cell *77*, 413–426.

Lu, Z., and Hunter, T. (2018). Metabolic kinases moonlighting as protein kinases. Trends Biochem. Sci. *43*, 301–310.

Ma, R., Wu, Y., Zhai, Y., Hu, B., Ma, W., Yang, W., Yu, Q., Chen, Z., Workman, J.L., Yu, X., et al. (2019). Exogenous pyruvate represses histone gene expression and inhibits cancer cell proliferation via the NAMPT–NAD+–SIRT1 pathway. Nucleic Acids Res. *47*, 11132–11150.

Maciag-Dorszynska, M., Ignatowska, M., Jannière, L., Wegrzyn, G., and Szalewska-Pałasz, A. (2012). Mutations in central carbon metabolism genes suppress defects in nucleoid position and cell division of replication mutants in *Escherichia coli*. Gene *503*, 31–35.

Maciąg, M., Nowicki, D., Jannière, L., Szalewska-Pałasz, A., and Wegrzyn, G. (2011). Genetic response to metabolic fluctuations: correlation between central carbon metabolism and DNA replication in *Escherichia coli*. Microb. Cell Fact. *10*, 19.

Magill, N.G., and Setlow, P. (1992). Properties of purified sporlets produced by spoII mutants of Bacillus subtilis. J. Bacteriol. *174*, 8148–8151.

Mathews, C.K. (2015). Deoxyribonucleotide metabolism, mutagenesis and cancer. Nat. Rev. Cancer *15*, 528–539.

Maya-Mendoza, A., Moudry, P., Merchut-Maya, J.M., Lee, M., Strauss, R., and Bartek, J. (2018). High speed of fork progression induces DNA replication stress and genomic instability. Nature *559*, 279–284.

Mäder, U., Schmeisky, A.G., Flórez, L.A., and Stülke, J. (2012). SubtiWiki--a comprehensive community resource for the model organism Bacillus subtilis. Nucleic Acids Res. *40*, D1278–D1287.

Misra, S.K., Milohanic, E., Aké, F., Mijakovic, I., Deutscher, J., Monnet, V., and Henry, C. (2011). Analysis of the serine/threonine/tyrosine phosphoproteome of the pathogenic bacterium Listeria monocytogenes reveals phosphorylated proteins related to virulence. Proteomics *11*, 4155–4165.

Morigen, Odsbu, I., and Skarstad, K. (2009). Growth rate dependent numbers of SeqA structures organize the multiple replication forks in rapidly growing *Escherichia coli*. Genes to Cells *14*, 643–657.

Murray, H., and Koh, A. (2014). Multiple regulatory systems coordinate DNA replication with cell growth in *Bacillus subtilis*. PLoS Genet. *10*, e1004731.

Nguyen, C.C., and Saier, M.H., Jr (1995). Phylogenetic analysis of the putative phosphorylation domain in the pyruvate kinase of *Bacillus stearothermophilus*. Res. Microbiol. *146*, 713–719.

Nicolas, P., Mäder, U., Dervyn, E., Rochat, T., Leduc, A., Pigeonneau, N., Bidnenko, E., Marchadier, E., Hoebeke, M., Aymerich, S., et al. (2012). Condition-dependent transcriptome reveals high-level regulatory architecture in *Bacillus subtilis*. Science *335*, 1103–1106.

Noirot-Gros, M.-F., Dervyn, E., Wu, L.J., Mervelet, P., Errington, J., Ehrlich, S.D., and Noirot, P. (2002). An expanded view of bacterial DNA replication. Proc. Natl. Aca. Sci. USA *99*, 8342–8347.

Nouri, H., Monnier, A.-F., Fossum-Raunehaug, S., Maciag-Dorszynska, M., Cabin-Flaman, A., Képès, F., Wegrzyn, G., Szalewska-Palasz, A., Norris, V., Skarstad, K., et al. (2018). Multiple links connect central carbon metabolism to DNA replication initiation and elongation in *Bacillus subtilis*. DNA Res. *25*, 641–653.

Odsbu, I., Morigen, and Skarstad, K. (2009). A reduction in ribonucleotide reductase activity slows down the chromosome replication fork but does not change its localization. PLoS ONE *4*, e7617–13.

Oskam, L., Hillenga, D.J., Venema, G., and Bron, S. (1991). The large *Bacillus* plasmid pTB19 contains two integrated rolling-circle plasmids carrying mobilization functions. Plasmid *26*, 30–39.

Pancholi, V., and Chhatwal, G.S. (2003). Housekeeping enzymes as virulence factors for pathogens. Int. J. Med. Microbiol. *293*, 391–401.

Papagiannakis, A., Niebel, B., Wit, E.C., and Heinemann, M. (2017). Autonomous metabolic oscillations robustly gate the early and late cell cycle. Mol. Cell *65*, 285–295.

Paschalis, V., Le Chatelier, E., Green, M., Képès, F., Soultanas, P., and Jannière, L. (2017). Interactions of the *Bacillus subtilis* DnaE polymerase with replisomal proteins modulate its activity and fidelity. Open Biol. *7*, 170146.

Pisithkul, T., Patel, N.M., and Amador-Noguez, D. (2015). Post-translational modifications as key regulators of bacterial metabolic fluxes. Curr. Opin. Microbiol. *24*, 29–37.

Popanda, O., Fox, G., and Thielmann, H.W. (1998). Modulation of DNA polymerases alpha, delta and epsilon by lactate dehydrogenase and 3-phosphoglycerat kinase. Biochim. Biophys. Acta *1397*, 102–117.

Prakasam, G., Iqbal, M.A., Bamezai, R.N.K., and Mazurek, S. (2018). Posttranslational modifications of pyruvate kinase M2: Tweaks that benefit cancer. Front. Oncol. *8*, 309–312.

Rannou, O., Le Chatelier, E., Larson, M.A., Nouri, H., Dalmais, B., Laughton, C., Jannière, L., and Soultanas, P. (2013). Functional interplay of DnaE polymerase, DnaG primase and DnaC helicase within a ternary complex, and primase to polymerase hand-off during lagging strand DNA replication in *Bacillus subtilis*. Nucleic Acids Res. *41*, 5303–5320.

Reiland, S., Messerli, G., Baerenfaller, K., Gerrits, B., Endler, A., Grossmann, J., Gruissem, W., and Baginsky, S. (2009). Large-scale Arabidopsis phosphoproteome profiling reveals novel chloroplast kinase substrates and phosphorylation networks. Plant Physiol. *150*, 889–903.

Ronai, Z. (1993). Glycolytic enzymes as DNA binding proteins. Int. J. Biochem. Cell Biol. *25*, 1073–1076.

Sakai, H. (2004). Possible structure and function of the extra C-terminal sequence of pyruvate kinase from *Bacillus stearothermophilus*. J. Biochem. *136*, 471–476.

Sanders, G.M., Dallmann, H.G., and McHenry, C.S. (2010). Reconstitution of the *B. subtilis* replisome with 13 proteins including two distinct replicases. Mol. Cell *37*, 273–281.

Schaechter, M., Maaloe, O., and O, K.N. (1958). Dependency on medium and temperature of cell size and chemical composition during balanced growth of *Salmonella typhimurium*. J. Gen. Microbiol. *19*, 592–606.

Schormann, N., Hayden, K.L., Lee, P., Banerjee, S., and Chattopadhyay, D. (2019). An overview of structure, function, and regulation of pyruvate kinases. Protein Sci. *28*, 1771–1784.

Séror, S.J., Casarégola, S., Vannier, F., Zouari, N., Dahl, M., and Boye, E. (1994). A mutant cysteinyl-tRNA synthetase affecting timing of chromosomal replication initiation in *B. subtilis* and conferring resistance to a protein kinase C inhibitor. EMBO J. *13*, 2472–2480.

Sharpe, M.E., Hauser, P.M., Sharpe, R.G., and Errington, J. (1998). *Bacillus subtilis* cell cycle as studied by fluorescence microscopy: constancy of cell length at initiation of DNA replication and evidence for active nucleoid partitioning. J. Bacteriol. *180*, 547–555.

Sirover, M.A. (1999). New insights into an old protein: The functional diversity of mammalian glyceraldehyde-3-phosphate dehydrogenase. Biochim. Biophys. Acta, Gen. Subj. *1432*, 159–184.

Sirover, M.A. (2011). On the functional diversity of glyceraldehyde-3-phosphate dehydrogenase: Biochemical mechanisms and regulatory control. Biochim. Biophys. Acta *1810*, 741–751.

Skarstad, K., Lobner-Olesen, A., Atlung, T., Meyenburg, von, K., and Boye, E. (1989). Initiation of DNA replication in *Escherichia coli* after overproduction of the DnaA protein. Mol. Gen. Genet. *218*, 50–556.

Snaebjornsson, M.T., and Schulze, A. (2018). Non-canonical functions of enzymes facilitate cross-talk between cell metabolic and regulatory pathways. Exp. Mol. Med. *50*, 34.

Soultanas, P. (2012). Loading mechanisms of ring helicases at replication origins. Mol. Microbiol. *84*, 6–16.

Sprague, G.F.J. (1977). Isolation and characterization of a *Saccharomyces cerevisiae* mutant deficient in pyruvate kinase activity. J. Bacteriol. *130*, 232–241.

Stein, A., and Firshein, W. (2000). Probable identification of a membrane-associated repressor of *Bacillus subtilis* DNA replication as the E2 subunit of the pyruvate dehydrogenase complex. J. Bacteriol. *182*, 2119–2124.

Sutendra, G., Kinnaird, A., Dromparis, P., Paulin, R., Stenson, T.H., Haromy, A., Hashimoto, K., Zhang, N., Flaim, E., and Michelakis, E.D. (2014). A nuclear pyruvate dehydrogenase complex Is important for the generation of Acetyl-CoA and histone acetylation. Cell *158*, 84–97.

Suzuki, K., Ito, S., Shimizu-Ibuka, A., and Sakai, H. (2008). Crystal structure of pyruvate kinase from *Geobacillus stearothermophilus*. J. Biochem. *144*, 305–312.

Teplyakov, A., Lim, K., Zhu, P.P., Kapadia, G., Chen, C.C.H., Schwartz, J., Howard, A., Reddy, P.T., Peterkofsky, A., and Herzberg, O. (2006). Structure of phosphorylated enzyme I, the phosphoenolpyruvate:sugar phosphotransferase system sugar translocation signal protein. Proc. Natl. Aca. Sci. USA *103*, 16218–16223.

Tolentino, R., Chastain, C., and Burnell, J. (2013). Identification of the amino acid involved in the regulation of bacterial pyruvate, orthophosphate dikinase and phosphoenolpyruvate synthetase. Advances in Biological Chemistry *03*, 12–21.

Tu, B.P., Kudlicki, A., Rowicka, M., and McKnight, S.L. (2005). Logic of the yeast metabolic cycle: temporal compartmentalization of cellular processes. Science *310*, 1152–1158.

Tymecka-Mulik, J., Boss, L., Maciag-Dorszynska, M., Matias Rodrigues, J.F., Gaffke, L., Wosinski, A., Cech, G.M., Szalewska-Pałasz, A., Wegrzyn, G., and Glinkowska, M. (2017). Suppression of the *Escherichia coli dnaA46* mutation by changes in the activities of the pyruvate-acetate node links DNA replication regulation to central carbon metabolism. PLoS ONE *12*, e0176050.

Wang, J.D., and Levin, P.A. (2009). Metabolism, cell growth and the bacterial cell cycle. Nat. Rev. Microbiol. *7*, 822–827.

Washington, T.A., Smith, J.L., and Grossman, A.D. (2017). Genetic networks controlled by the bacterial replication initiator and transcription factor DnaA in *Bacillus subtilis*. Mol. Microbiol. *106*, 109–128.

Wellen, Hatzivassililiou, G., Sachdeva, U.M., Bui, T.V., Cross, J.R., and Thompson, C.B. (2009). ATP-citrate lyase links cellular metabolism to histone acetylation. Science *324*, 1071–1076.

Wieczorek, A., Fornalewicz, K., Mocarski, Ł., Łyżeń, R., and Wegrzyn, G. (2018). Double silencing of relevant genes suggests the existence of the direct link between DNA replication/repair and central carbon metabolism in human fibroblasts. Gene *650*, 1–6.

Yu, F.-X., Dai, R.-P., Goh, S.-R., Zheng, L., and Luo, Y. (2009). Logic of a mammalian metabolic cycle: an oscillated NAD+/NADH redox signaling regulates coordinated histone expression and S-phase progression. Cell Cycle *8*, 773–779.

Zhang, Q., Zhou, A., Li, S., Ni, J., Tao, J., Lu, J., Wan, B., Li, S., Zhang, J., Zhao, S., et al. (2016). Reversible lysine acetylation is involved in DNA replication initiation by regulating activities of initiator DnaA in *Escherichia coli*. Sci. Rep. *6*, 30837.

Zheng, L., Roeder, R.G., and Luo, Y. (2003). S phase activation of the histone H2B promoter by OCA-S, a coactivator complex that contains GAPDH as a key component. Cell *114*, 255–266.